\documentclass[twocolumn,aps,showpacs,pre,superscriptaddress]{revtex4}

\usepackage{graphicx}
\usepackage{amssymb}
\usepackage{latexsym}

\newcommand{\BEQ}{\begin{equation}} 
\newcommand{\EEQ}{\end{equation}} 
\newcommand{\BEA}{\begin{eqnarray}} 
\newcommand{\EEA}{\end{eqnarray}} 
\newcommand{\nn}{\nonumber \\} 
\renewcommand{\d}{{\rm d}} 
 
\newcommand{\oq}{\frac{1}{4}\,} 
\newcommand{\os}{\frac{1}{6}\,}

\newcommand{\mt}{{\bf m}} 
\newcommand{\eps}{\,\epsilon\,} 
\newcommand{\epseff}{\,\epsilon_{\rm eff}\,} 
\newcommand{\al}{\alpha} 
\newcommand{\va}{v_{\rm b}} 
\newcommand{\rt}{{\bf {\rho}}}

\newcommand{\qb}{{\bf q}}

\newcommand{\mb}{{\bf m}} 
\newcommand{\nulb}{{\bf 0}} 
\newcommand{\half}{\frac{1}{2}\,} 
\newcommand{\bd}{{\rm b}} 
\newcommand{\ub}{{\rm ub}} 
\newcommand{\taun}{{\tau_0}} 
\begin{document}

\title{Random walks of molecular motors arising from
diffusional encounters with immobilized filaments}
\author{Theo M.\ Nieuwenhuizen}\email{nieuwenh@science.uva.nl}
\affiliation{Institute for Theoretical Physics, University of Amsterdam, Valckenierstraat 65, 1018 XE Amsterdam, The Netherlands}
\affiliation{Max Planck Institute of Colloids and Interfaces, 14424 Potsdam, Germany}
\author{Stefan Klumpp}\email{klumpp@mpikg-golm.mpg.de}
\author{Reinhard Lipowsky}\email{lipowsky@mpikg-golm.mpg.de} 
\affiliation{Max Planck Institute of Colloids and Interfaces, 14424 Potsdam, Germany}
\date{\today}

\begin{abstract} 
Movements of molecular motors on cytoskeletal filaments are described by  
directed walks on a line. Detachment from this line is 
allowed to occur with a small probability. Motion in the  
surrounding fluid is described by symmetric random walks.  
Effects of detachment and reattachment are calculated by an analytical  
solution of the master equation in two and three dimensions. 
Results are obtained for the fraction of bound motors, their 
average velocity and displacement. The diffusion coefficient 
parallel to the filament becomes anomalously large since
detachment and  subsequent reattachment, in the presence 
of directed motion of the bound motors,  leads to a broadening 
of the density distribution.

The occurrence of protofilaments on a microtubule is modeled  
by internal states of the binding sites.  
After a transient time all protofilaments become equally populated. 
\end{abstract} 

 
\maketitle


\section{Introduction} 
 
Molecular motors are proteins that convert the free energy released 
from chemical reactions into directed movements 
\cite{Schliwa2003,Howard2001}. Here we focus on linear cytoskeletal 
motors which move along cytoskeletal filaments. Many of these motors 
are processive in the sense that a single motor molecule can move a 
cargo over a large distance. The most prominent examples are 
(conventional) kinesin and certain types of myosins, which move along 
microtubules and actin filaments, respectively. In the cell, these 
motors are involved in transport processes, reorganization of the 
cytoskeleton and cell division \cite{Schliwa2003}.  However, 
experiments on the movements of molecular motors can also be done in 
vitro, which has lead to the development of various single molecule 
assays. In these experiments, one can measure the velocities, step 
sizes, walking distances and forces for single motor molecules, see, 
e.g.\ \cite{Howard2001}. In additions, they have stimulated a lot of 
theoretical work devoted to the walks of molecular motors along 
filaments, see, e.g.\ \cite{Juelicher__Prost1997}. 
 
In the experiments, one observes that even processive motors unbind 
 from their filamentous tracks after a certain walking distance, which 
is typically of the order of 1~$\mu$m. For a kinesin molecule, this 
means that it makes about 100 steps of 8~nm before unbinding 
\cite{Block__Schnapp1990,Vale__Yanagida1996}. Myosin~V motors have a 
comparable walking distance, but a larger step size of 36~nm, so that 
they detach after about 30--50 steps 
\cite{Mehta__Cheney1999,Veigel__Molloy2002}.  Unbound motors then 
diffuse in the surrounding fluid until they rebind to the same or 
another filament and continue their directed walk. 
 
\begin{figure}[h] 
  \begin{center} 
    \leavevmode 
    \includegraphics[angle=0,width=.5\textwidth]{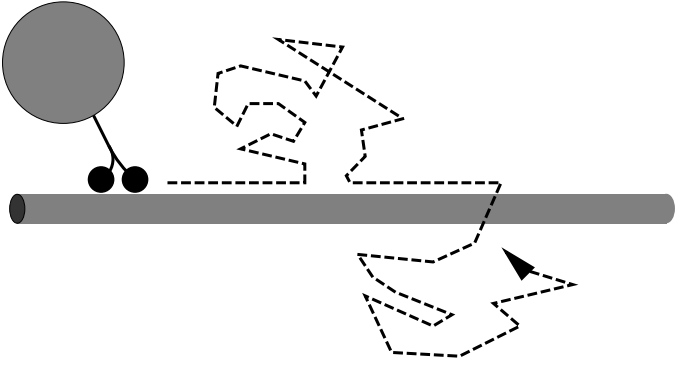}
    \caption{Random walk of a molecular motor: The motor performs directed  
      movement along a filament (grey rod) and unbinds from it after a 
      certain walking distance. The unbound motor diffuses in 
      the surrounding fluid until it rebinds to the filament and 
      resumes directed motion.} 
    \label{fig:randWalk} 
  \end{center} 
\end{figure}

On larger scales, the motors thus perform complex random walks, which 
consist of alternating sequences of directed movements along filaments 
and non-directed diffusion in the surrounding fluid, as shown 
schematically in Fig.\ \ref{fig:randWalk}. These random walks have 
been discussed by Ajdari using scaling arguments \cite{Ajdari1995}. 
Recently, we have introduced lattice models to describe the random 
walks of the motors as random walks on a lattice, where certain lines 
of lattice sites represent the filaments 
\cite{Lipowsky__Nieuwenhuizen2001,Nieuwenhuizen__Lipowsky2002}. When 
bound to these lines, the motors perform directed random walks. 
Detachment from these lines is allowed to occur with a small, but 
non-zero probability. Diffusive motion in the surrounding fluid is 
described by symmetric random walks. 
 
These models are designed to study generic properties of motor 
movements, but they can also be used to describe specific motor 
molecules, since all model parameters can be determined from the 
measured transport properties, see Ref.\ \cite{Lipowsky__Nieuwenhuizen2001}. 
In addition, motor--motor 
interactions can be easily included into these models, e.g., mutual 
exclusion of motors from the binding sites of the filaments, which 
leads to self-organized density profiles in closed systems 
\cite{Lipowsky__Nieuwenhuizen2001} and boundary-induced phase 
transitions in open tube systems \cite{Klumpp_Lipowsky2003}. 
 
For the random walks of single motors or, equivalently, for an 
ensemble of non-interacting motors, we have obtained a number of exact 
results for the cases of a single filament embedded in two-dimensional 
or three-dimensional space as reported in 
\cite{Nieuwenhuizen__Lipowsky2002}. In particular, these random walks 
exhibit anomalous drift behavior; the average position of the motor 
advances slower than linearly with time.  The same drift behavior is 
found for the movement along a single filament immobilized in open 
compartments with the same dimensionality, which are more easily 
accessible to experiments \cite{Lipowsky__Nieuwenhuizen2001}. In the 
present article, we present a detailed derivation of the analytical 
results of Ref.\ \cite{Nieuwenhuizen__Lipowsky2002} for movements in 
two and three dimensions without confining boundaries. 
 
Analytical results are obtained by the following method, which is a 
variant of the method of Fourier--Laplace transforms for random walks 
in homogeneous space, see, e.g., \cite{Weiss1994,Haus_Kehr1986}: By 
using Fourier--Laplace transforms of the probability distributions, 
the master equations of the random walk can be transformed into a set 
of algebraic equations, one of which, however, requires the evaluation 
of a non-trivial integral. Solving these algebraic equations, 
solutions for the Fourier--Laplace transformed probability 
distributions and their moments are obtained and closed expressions in 
terms of integrals can be derived for the time-dependent probability 
distributions and moments. These can, on the one hand, be evaluated 
numerically to obtain results for all times; on the other hand, 
asymptotic results for small and large times can be obtained fully 
analytically by using the Tauberian theorems, which we summarize in 
the appendix.  In this way, we derive expressions for the fraction of 
bound motors, the average displacement and dispersion, and the 
effective velocities and diffusion coefficients.  The analytical 
results are compared to data from Monte Carlo simulation and are found 
to be in very good agreement. 
 
In addition to the anomalous drift behavior, the random walks of 
molecular motors also exhibit strongly enhanced diffusion in the 
direction parallel to the filament. 
  
Our article is organized as follows: We start with the two-dimensional 
case in section \ref{sec:2d} and discuss the three-dimensional case in 
section \ref{sec:3d}. In both cases, we derive probability 
distributions and their moments for both the bound and unbound motor. 
The fact that filaments may consist of several protofilaments is taken 
into account in the final subsections of sections \ref{sec:2d} and 
\ref{sec:3d}, where these are modeled by several internal states of 
the bound motors. In section \ref{sec:sticking_prob}, we extend the 
discussion to include a variable sticking probability for motors 
arriving at the filament. At the end, we include a short summary of 
our results.

\section{Random walks in two dimensions} 
\label{sec:2d} 
 
\begin{figure}[h] 
  \begin{center} 
    \leavevmode
    \includegraphics[angle=0,width=.5\textwidth]{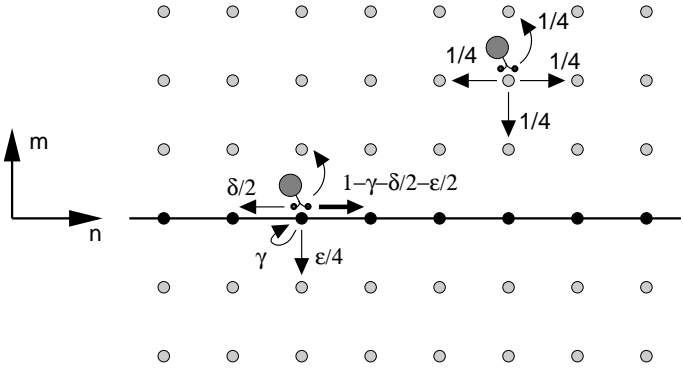}
    \caption{The random walks of molecular motors are modeled as  
      random walks on a lattice. A line of lattice sites, which is  
      indicated here as a black line, represents a  
      filament. Motors at filament sites perform a directed random walks,  
      while motors at non-filament sites undergo symmetric random walks.  
      For the movement in two dimensions, the jump probabilities at  
      non-filament sites are $1/4$ for each of the four neighbor sites,  
      at filament sites, a motor steps forward with probability  
      $1-\gamma-\delta/2-\epsilon/2$ and backward with probability  
      $\delta/2$; jumps to each of the adjacent non-filament sites  
      which lead to unbinding occur with probability $\epsilon/4$.
      The dwell probability is $\gamma$.} 
    \label{fig:randWalk_2d} 
  \end{center} 
\end{figure}

Consider a discrete time random walk on a two dimensional square 
lattice with lattice sites labeled by integer coordinates $(n,m)$.  At 
each step a particle has probability 1/4 to jump into any of the four 
directions.  For modeling the motion of a motor on a microtubule, we 
choose different behavior on the line with $m=0$.  Here the 
probability to jump from $(n,0)$ to $(n,\pm 1)$ equals 
$\frac{1}{4}\epsilon$, while jumps to $(n+1,0)$ have probability 
$1-\gamma-\frac{1}{2}\delta-\frac{1}{2}\epsilon$; the probability to 
jump to $(n-1,0)$ is $\frac{1}{2}\delta$ and the probability to make 
no jump is $\gamma$, see Fig.\ \ref{fig:randWalk_2d}.  
The latter parameter is needed for modeling realistic 
situations, where the diffusion coefficient in the fluid is much 
larger than on the filament \cite{Lipowsky__Nieuwenhuizen2001}.  The 
ordinary random walk in two dimensions has $\gamma=0$, $\delta=1/2$, 
and $\epsilon=1$.  We shall assume that the escape probability 
$\epsilon$ is small.  For $\epsilon=0$ the problem amounts to a 
directed random walk on the line with $m=0$. The average speed of a 
motor particle on the filament line is 
$\va =1-\gamma-\delta-\half\, \epsilon$.  Per step there is a 
probability $\half\eps$ to unbind.  The probability that the motor is 
still bound after $t$ steps is 
\begin{equation} \label{punb=} 
(1-\half\eps)^t\approx \exp(-\half \eps\, t). \end{equation} 
 
The master equation for this dynamics reads 
\begin{widetext}
\begin{eqnarray} \label{me1} 
P_{n,m}(t+1)&=&\oq P_{n+1,m}+\oq P_{n-1,m}+\oq P_{n,m+1}+\oq P_{n,m-1} 
\qquad(m\neq 0,\pm 1)\\ \label{me2} 
P_{n,0}(t+1)&=&\oq P_{n,1}+\oq P_{n,-1}+ 
(1-\gamma-\half\epsilon-\half\delta)P_{n-1,0} 
+\frac{\delta}{2} P_{n+1,0}+\gamma P_{n,0}\\ \label{me3} 
P_{n,1}(t+1)&=&\oq P_{n+1,1}+\oq P_{n-1,1}+\oq P_{n,2}+ 
\frac{\epsilon}{4} P_{n,0}\\ \label{me4} 
P_{n,-1}(t+1)&=&\oq P_{n+1,-1}+\oq P_{n-1,-1}+\oq P_{n,-2} 
+\frac{\epsilon}{4} P_{n,0}. \label{2dmeq} 
\end{eqnarray} 
\end{widetext} 
As initial condition we take an ensemble of particles at $n=m=0$, so 
\begin{equation}  
  P_{n,m}(t=0)=\delta_{n,0}\delta_{m,0}.
\end{equation} 
Let us now define the Fourier--Laplace transforms of the probability 
distribution along the filament $P_\bd(n,t)\equiv P_{n,0}(t)$ and the 
full distribution $P_{n,m}(t)$, 
\begin{equation}\label{p0rz}  
  P_\bd(r,s)\equiv \sum_{t=0}^\infty\sum_{n=-\infty}^\infty 
  \frac{e^{ir n}}{(1+s)^{t+1}} P_{n,0}(t) 
\end{equation} 
and 
\begin{equation}  
  P(q,r,s)\equiv \sum_{t=0}^\infty\sum_{m,n=-\infty}^\infty 
  \frac{e^{iq m+ir n}}{(1+s)^{t+1}} P_{n,m}(t). 
\end{equation}  
Then the master equations are reduced to an algebraic equation 
relating $P_\bd(r,s)$ and $P(q,r,s)$ as given by 
\begin{widetext}
\begin{eqnarray} \label{p1a=} 
  \lefteqn{(1+s-\half \cos q-\half\cos r)P(q,r,s)=1  }\nonumber\\ 
  & & +\left[\left( [1-\gamma-\frac{\delta}{2}-\frac{\epsilon}{2}]-\frac{1}{4}\right)e^{ir}+\left(\frac{\delta}{2}-\frac{1}{4}\right)e^{-ir} 
  +\gamma-\frac{1-\epsilon}{2}\cos q\right]\, P_\bd(r,s). 
\end{eqnarray} 
Here the first line is what one would get in the case of a symmetric 
random walk in two dimensions, and the second line corrects those 
terms that are changed by the presence of the filament. This equation 
has the obvious solution 
\begin{eqnarray} \label{p2=} 
  P(q,r,s)=\frac{1+[\gamma(1-\cos r) 
    +\half(1-\eps)(\cos r-\cos q)+i\va \sin r]P_\bd(r,s) } 
  {s+1-\half\cos q-\half\cos r}. 
\end{eqnarray} 
By integrating this result over $q$ we also obtain $P_\bd(r,s)$ on the 
left hand side. It thus satisfies a linear equation, that can be 
easily solved. Introducing the variable $\mu$ via 
\begin{equation}  
  \cosh\mu  \equiv  2+2s-\cos r\qquad
  \rm{or}\qquad \sinh\mu  = \sqrt{(2+2s-\cos r)^2-1},  
\end{equation} 
we may use the equalities 
\begin{equation}
  \int_0^{2\pi} \frac{\d q}{2\pi}\frac{1}{\cosh\mu-\cos q}  =  \frac{1}{\sinh\mu}\qquad
  \rm{and}\qquad \int_0^{2\pi} \frac{\d q}{2\pi}\frac{\cos q}{\cosh\mu-\cos q} 
 = \frac{e^{-\mu}}{\sinh\mu}. \label{Imu=}   
\end{equation} 
After some computation, we then end up with the probability 
distribution 
\begin{eqnarray}  
  P_\bd(r,s)&=&\frac{1}{s+1-\gamma-(1-\gamma-\half\delta-\half\epsilon) 
e^{ir} 
-\half\delta e^{-ir}-\half\epsilon\, e^{-\mu}}\nn 
&=&\frac{1}{s+(1-\gamma)(1-\cos r) 
+\half\eps(\cos r-e^{-\mu})-i\va\sin r} 
 \label{sold=2} 
\end{eqnarray}
 \end{widetext}
for the motors bound to the filaments. The probability distribution 
for all, bound and unbound motors follows via Eq.\ (\ref{p2=}).  It is 
easy to check that this is correct for $\epsilon=0$ (random walk in 
one dimension) and for $\gamma=0$, $\delta=\half$, and $\epsilon=1$ 
(non-biased random walk in 2d).

\subsection{Properties of the motors bound to the line} 
 
\subsubsection{Survival fraction} 
 
\begin{figure}[h] 
  \begin{center} 
    \leavevmode 
    \includegraphics[angle=-90,width=.5\textwidth]{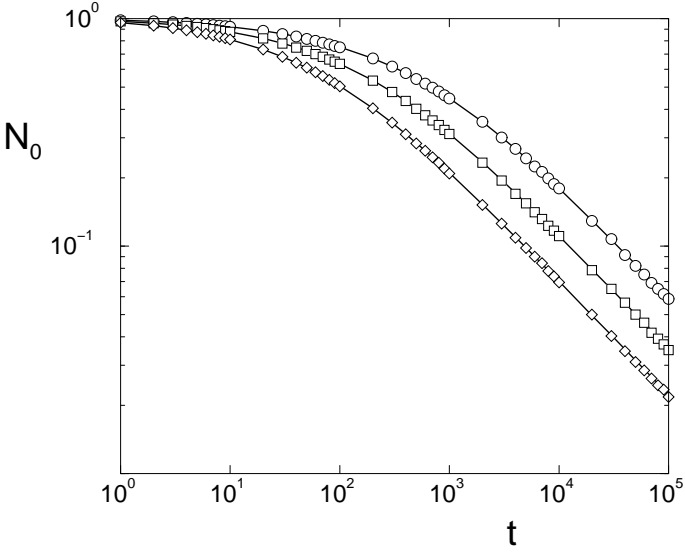}
    \caption{Fraction of motors bound to the filament, $N_0$, as a  
      function of time $t$ for the two-dimensional case. The three 
      curves correspond to $\epsilon=0.03$ (circles), $\epsilon=0.05$ 
      (squares) and $\epsilon=0.08$ (diamonds). Lines are obtained 
      from the exact integral (\ref{Nnult=}), the data points from 
      Monte Carlo simulation.  In the simulations, the other 
      parameters have been chosen as $\gamma=0$, $\delta=0.6$, 
      but the results shown here depend only on $\epsilon$.} 
    \label{fig:Pb_d2_eps.03.05.08} 
  \end{center} 
\end{figure}

In the following, we extract the transport properties of the motor's 
random walks from the solution (\ref{sold=2}).  The value at $r=0$ 
gives us the Laplace transform $N_0(s)$ of the probability 
$N_0(t)\equiv \sum_n P_{n,0}(t)$ that the motor particle is bound to 
the filament line with $m=0$: 
\begin{widetext}
\begin{equation} \label{N0s=} 
  N_0(s)=\sum_{t=0}^\infty \frac{N_0(t)}{(1+s)^{t+1}} 
  =\frac{1}{s+\frac{1}{2}\epsilon(1-e^{-\mu})} 
  =\frac{1}{(1-\eps)s+\eps\sqrt{s(1+s)}} 
\end{equation} 
The inverse is 
\begin{equation} \label{Nnult=} 
  N_0(t)=\oint\frac{\d s}{2\pi i}(1+s)^{t}\,N_0(s) 
  =\int_0^1\frac{\d x}{\pi}\frac{\eps (1-x)^{t+1/2}}{\sqrt{x}\, 
    [\eps^2+(1-2\eps)x]} 
  =\int_0^{\eps^{-2}}\frac{\d y}{\pi}\frac{(1-\eps^2 y)^{t+1/2}}{\sqrt{y}\, 
    [1+(1-2\eps)y]}. 
\end{equation} 
\end{widetext}
This expression is exact. It holds for all $t$ and for all 
$\epsilon$, $\delta$ and can be evaluated numerically. Values for 
$N_0(t)$ obtained in this way are shown Fig.\  
\ref{fig:Pb_d2_eps.03.05.08} (lines) for three different values of the 
detachment rate $\epsilon$.  Comparison to results of Monte Carlo 
simulations (data points) shows that the agreement is very good. 
 
Let us now derive the asymptotic behavior for small and large times, 
respectively.  If $\eps$ is small and $t\gg 1$, we have 
\begin{equation} \label{N0s2} 
  N_0(s)\approx \frac{1}{s+\epsilon \sqrt{s}}. 
\end{equation} 
Then the inverse behaves as 
\begin{equation} \label{N0t=} 
  N_0(t)\approx\int_{-i\infty}^{i\infty}\frac{\d s}{2\pi i} 
  \,\frac{e^{st}}{s+\epsilon\sqrt{s}} 
  =\int_0^\infty\frac{\d y}{\pi\sqrt{y}}\,\,\frac{e^{-y\epsilon^2t}} 
  {y+1}. 
\end{equation} 
The second integral is an obvious limit of the last integral in Eq.\  
(\ref{Nnult=}).  For deriving the short-time behavior we may start 
 from the series of equalities as given by 
\begin{widetext}
\begin{eqnarray}  
  \frac{e^{-y\epsilon^2t}}{y+1} 
  &=&\frac{1}{y+1}+\frac{e^{-y\epsilon^2t}-1}{y+1} 
  =\frac{1}{y+1}+\frac{e^{-y\epsilon^2t}-1}{y}+ 
  \frac{\epsilon^2t}{y+1}-\frac{e^{-y\epsilon^2t}+y\epsilon^2t-1}{y^2} 
  +\cdots 
\end{eqnarray} 
The integrals over the exponential terms are most easily carried out 
using dimensional regularization. To show how that works, let us 
consider the last term.  We need to consider the expression 
\begin{equation} 
  -\frac{1}{\pi}\int_0^\infty\d y y^{n-5/2}(e^{-y\epsilon^2t}+y\epsilon^2t-1) 
  =-\frac{(\epsilon^2t)^{3/2-n}}{\pi}\Gamma(n-\frac{3}{2})= 
  -\frac{(\epsilon^2t)^{3/2-n}}{\pi}\, 
  \frac{\Gamma(n+\half)}{(n-\frac{3}{2})(n-\half)} 
\end{equation} 
\end{widetext}
in the limit of small $n$, where we used that, in dimensional 
regularization, integrals of powers are put equal to zero. The limit 
$n\to 0$ can now be taken. Using the same procedure for the other 
terms, we obtain 
\begin{equation} \label{N0t<=} 
  N_0(t)\approx 1-2\,\frac{\epsilon \sqrt t}{\sqrt\pi}+\epsilon^2t 
  -\frac{4\epsilon^3t^{3/2}}{3\sqrt\pi}+\cdots 
\end{equation} 
for small $t$, which represents a series in powers of $\epsilon\sqrt 
t$.  For $t\ll 1/\epsilon^2$, this is somewhat surprising: although 
the motors detach at times $\sim 1/\eps$, the recurrent behavior of 
the random walk brings them mostly back to the filament.  For $t\ll 
1/\epsilon^2$, this just says that the motor did not have enough time 
to escape from the line.  The half-integer powers are related to the 
long range diffusion away from the filament. 
 
Note that, for the short time limit, we have assumed that 
$t\gg 1 $, i.e., that $t$ is large compared to the time 
required for one step of the random walk. 
In that interval the approximation (\ref{N0s2}) holds with the term 
$\eps\sqrt{s}$ being due to diffusion. 
 
For smaller times, the 
random walk exhibits discrete steps. Our short time result would hold 
for arbitrarily small times in the limit in which the walk on the line 
becomes a continuous time random walk. A continuous time random walk 
with exponential waiting time distribution is obtained approximately, 
if $1-\gamma$ is small, which is the case for realistic applications 
of our model to molecular motor setups 
\cite{Lipowsky__Nieuwenhuizen2001}. The same remark holds for all the 
short time results discussed in the following.

Another way to derive (\ref{N0t<=}) is to expand (\ref{N0s=}) in 
powers of $1/\sqrt s$ and to use the Tauberian theorem which states 
that the inverse Laplace transform of $N_0(s) = as^{-\alpha}$ is given 
by 
\begin{equation} \label{Tauber} 
  N_0(t)=\frac{a}{\Gamma(\alpha)}t^{\alpha-1}, 
\end{equation} 
see appendix \ref{sec:app:TaubThms} and, e.g., Ref.\  
\cite{Feller1971}.  This theorem holds both for positive and negative 
values of $\alpha$.  It also shows that positive integer powers of $s$ 
of $N_0(s)$ in the limit of small $s$ do not contribute to long time 
tails. 
 
For large times, $t\gg 1/\epsilon^2$, the expression (\ref{N0t=}) for 
$N_0(t)$ can be evaluated using the expansion $1/(1+y)\approx 1-y+y^2$ 
for small $y$ which leads to 
\begin{equation} \label{N0t>=} 
  N_0(t)\approx  
  \frac{1}{\sqrt{\pi\,\epsilon^2\,t}}(1-\frac{1}{2\epsilon^2t} 
  +\frac{3}{4\epsilon^4t^2}). 
\end{equation} 
Alternatively, one can expand (\ref{N0s=}) in powers of $\sqrt{s}$ and 
use again the Tauberian theorem. It follows from (\ref{N0t>=}) that 
for large $t$ the probability to be bound to the filament decays as 
$t^{-1/2}$ in agreement with scaling arguments 
\cite{Ajdari1995,Lipowsky__Nieuwenhuizen2001}.

\subsubsection{Average position and speed on the filament line} 
 
The expression (\ref{sold=2}) for the Fourier--Laplace transformed 
probability distribution $P_\bd(r,s)$ contains much more information. 
At linear order in $r$ we get the Laplace transform of the average 
position of motor particles along the filament line, 
\begin{equation}  
N_1(t)\equiv\sum_{n}n P_{n,0}(t)=-i\frac{\partial}{\partial r}P_\bd(r,t)\Big|_{r=0}. 
\end{equation}  
We obtain 
\begin{equation}  
  N_1(s)=\va N_0^2(s)=\frac{\va} 
  {[(1-\epsilon)s+\epsilon\sqrt{s(1+s)}]^2}. 
\end{equation}  
In the limit of small $\epsilon$ and large $t$, this implies 
\begin{equation}  
  N_1(s)\approx\frac{\va } {[s+\epsilon\sqrt{s}]^2}. 
\end{equation}  
We invert the Laplace transform by taking the $1/s$ term apart.  We 
then get for the average position the asymptotic behavior 
\begin{eqnarray}  
  N_1(t) & \approx & \frac{\va }{\epsilon^2}-2\va \epsilon\int_0^\infty \frac{\d 
    x}{\pi \sqrt{x}} \,\frac{e^{-xt}}{(\epsilon^2+x)^2} \nn
   & = &\frac{2\va}{\pi 
    \epsilon^2}\int_0^\infty \frac{\d y}{\sqrt{y}} 
  \,\frac{1-e^{-y\epsilon^2t}}{(1+y)^2} \label{N1t=i}
\end{eqnarray}  
for large $t$.  The exact expression is 
\begin{equation} \label{N1t=i_exact} 
  N_1(t)=\frac{2\va(1-\eps)}{\pi\eps^2}\int_0^{\eps^{-2}}{\d 
    y}\,\frac{\sqrt{1-\eps^2 y}[1-(1-\eps^2 y)^{t}]}{ \sqrt{y}\,[ 1+y(1-2\eps)]^2} 
  , 
\end{equation}  
which deviates from (\ref{N1t=i}) for times of order unity.  The full 
expression (\ref{N1t=i_exact}) is evaluated numerically and plotted in 
Fig.\ \ref{fig:displ_d2_b_ub_all}. The same figure contains data 
points as obtained from Monte Carlo simulations which confirm the 
analytical result (\ref{N1t=i_exact}). 
 
For short times we proceed as above. We expand 
\begin{eqnarray}  
  N_1(t) & \approx &\frac{2\va }{\pi \epsilon^2}\int_0^\infty \frac{\d 
    x}{\sqrt{x}} \left[ 
    \frac{x\epsilon^2t}{(1+x)^2}+\,\frac{1-x\epsilon^2t 
      -e^{-x\epsilon^2t}}{x^2}\right]\nn
  & = & \va 
  t\,(1-\frac{8}{3}\,\frac{\epsilon \sqrt t}{\sqrt\pi}) 
\end{eqnarray}  
This leads to $N_1\approx\va t$. Thus, the average position 
$\bar n_\bd$ and speed $\bar v_\bd$ of the motors bound 
to the filament are given by 
\begin{eqnarray} 
  \bar n_\bd(t) &\equiv &\frac{N_1(t)}{N_0(t)}\approx \va 
  t(1-\frac{2}{3}\,\frac{\epsilon \sqrt t}{\sqrt\pi})\nn
  {\rm and}\qquad \bar v_\bd & \equiv &\frac{\d \bar n_\bd}{\d t} \approx\va (1-\frac{ \epsilon \sqrt 
    t}{\sqrt\pi}),  
\end{eqnarray}  
where $v_\bd$ is the average speed if the particles did not leave the 
line. 
 
For large times, the asymptotic expression (\ref{N1t=i}) leads to 
\begin{eqnarray}  
  N_1(t) & \approx &\frac{\va }{\epsilon^2}(1-\frac{2}{\epsilon\sqrt{\pi t}}) 
  \nn
  \rm{and}\qquad \bar n_\bd(t) &\approx &\frac{\va \sqrt{\pi t}}{\epsilon} 
  (1-\frac{2}{\epsilon\sqrt{\pi t}}).  
\end{eqnarray}  
Therefore, for large $t$, the average speed of bound motors behaves as 
\begin{equation} \label{venhd=2} 
  \bar v_\bd(t)\approx\frac{\va \sqrt{\pi}}{2 \epsilon\sqrt{t}} 
  =\frac{\pi}{2}N_0(t)\,\va . 
\end{equation} 
The last relation confirms the scaling $\bar v_\bd(t)\sim \va N_0(t)$, 
which has been used in the scaling approach 
\cite{Ajdari1995,Lipowsky__Nieuwenhuizen2001}.  The effective motor 
velocity is reduced by a factor $\sim N_0(t)$, i.e., by the 
probability that a motor is in the bound state. The relation 
$\bar v_\bd\sim N_0 v_\bd$ also applies to a simple two-state 
random walk, where motion is directed in one of the states only. In 
contrast to the simple two-state random walk, however, the probability 
$N_0$ is time-dependent here.  The factor $\pi/2$ in (\ref{venhd=2}) 
is solely due to the fact that only the bound motors are considered. 
We will show below, that this factor is absent, if all, bound and 
unbound, motors are considered.

\subsubsection{Dispersion and diffusion coefficient on the filament line} 
 
By expanding the expression (\ref{sold=2}) for the Fourier--Laplace 
transformed probability $P_\bd(r,s)$ up to second order in $r$, the 
second moment of the distribution of bound motors is found to behave 
as 
\begin{eqnarray}  
  \lefteqn{N_2(s) \equiv\sum_n n^2 P_{n,0}(s)}\nn
   & & = 2\va^2N_0^3(s)+(1-\gamma-\epsilon+\frac{\epsilon}{4\sqrt{s+s^2}})N_0^2(s). 
\end{eqnarray} 
The Laplace transform is again inverted by complex integration.  In 
the contour integral for $N_2(t)$, we may replace $e^{st}$ by 
$e^{st}-1$, since the subtracted integral vanishes, as can be seen by 
closing the contour of integration in the right half plane.

Closing the contour along the negative real axis, we obtain 
\begin{widetext}
\begin{eqnarray} \label{2ndMomentBound_LF} 
  N_2(t) & = & \frac{6\va ^2(1-\eps)^2}{\pi\epsilon^4}\int_0^{\eps^{-2}}\d x\, 
  \frac{[(1-x\epsilon^2)^t-1]\sqrt{1-\eps^2x}}{[1+(1-2\eps) x]^3\sqrt{x}}  
  - \frac{2\va ^2}{\pi\epsilon^4}\int_0^{\eps^{-2}}\d x \, 
  \frac{[(1-x\epsilon^2)^t-1](1-\eps^2x)^{3/2}}{[1+(1-2\eps) x]^3x^{3/2}} 
  \nonumber\\ 
  & & {}
  +\frac{2(1-\gamma-\eps)}{\pi\eps^2}\int_0^{\eps^{-2}}\d x\, 
  \frac{[1-(1-x\epsilon^2)^t]\sqrt{1-\eps^2x}}{[1+(1-2\eps) x]^2 \sqrt{x}}  
  \nonumber\\ 
  & & {}- \frac{1}{4\pi\eps^2}\int_0^{\eps^{-2}}\d x\,\frac{(1-2\eps^2x) 
    [(1-x\epsilon^2)^t-1+\eps^2xt][1-(1-2\eps)x]}{[1+(1-2\eps) x]^2 x^{3/2} 
    \sqrt{1-\eps^2x}} \nonumber\\ 
  & \approx &  \frac{2\va^2}{\pi\epsilon^4} \int_0^\infty\d x \, 
  \frac{(1-e^{-x\epsilon^2 t})(1-3x)}{(1+x)^3x^{3/2}} 
  +\frac{2(1-\gamma)}{\eps^2}\int_0^\infty\frac{\d x}{\pi}\, 
  \frac{1-e^{-x\eps^2 t}} {(1+x)^2\sqrt{x}}\nonumber\\ 
  & & {}
  + \frac{1}{4\eps^2}\int_0^\infty\frac{\d x}{\pi}\, 
  \frac{(1-e^{-x\eps^2 t})(1-x)} {(1+x)^2x\sqrt{x}}. 
\end{eqnarray} 
We have evaluated this exact expression numerically and compared it 
with simulation data. 
As shown in Fig.\ \ref{fig:deltanb2_d23}, the agreement is again very 
good. 
 
For short times we proceed as above. 
\begin{eqnarray}  
  N_2(t) &\approx&\frac{2\va ^2}{\pi\epsilon^4}\int_0^\infty\d x \, 
  \left[ 
    \frac{(x\epsilon^2 t-\half x^2\epsilon^4 t^2)(1-3x)}{(1+x)^3x^{3/2}} 
    +3\frac{e^{-x\epsilon^2 t}-1+x\epsilon^2 t-\half x^2\epsilon^4 t^2} 
    {x^{7/2}}\right]\nn 
  &=&\va ^2t^2 
  (1-\frac{16}{5}\,\frac{\epsilon \sqrt t}{\sqrt\pi})+ 
  (1-\gamma)t+\frac{8\gamma-7}{3}\,\frac{\epsilon t\sqrt t}{\sqrt\pi}. 
\end{eqnarray} 
\end{widetext}
This implies the normalized second moment 
\begin{equation}  
  \overline{n^2}_\bd\equiv\frac{N_2(t)}{N_0(t)} 
  \approx\va ^2t^2(1-\frac{6}{5}\,\frac{\epsilon \sqrt t}{\sqrt\pi}) 
  +(1-\gamma)t+\frac{2\gamma-1}{3}\,\frac{\epsilon t\sqrt t}{\sqrt\pi} 
\end{equation} 
and the dispersion 
\begin{equation}  
  \Delta n^2_\bd \equiv\overline{n^2}_\bd-\overline{n}^2_\bd\approx 
  \frac{2\va^2}{15}\,\frac{\epsilon\,t^2\sqrt{t}}{\sqrt\pi}\, 
  +(1-\gamma)t+\frac{2\gamma-1}{3}\,\frac{\epsilon t\sqrt t}{\sqrt\pi} 
\end{equation} 
This result holds again in the limit of continuous time, 
i.e.\ for $\gamma$ close to one. The leading term taking into account 
the time discretization is $(1-\gamma-v_\bd^2) t$, i.e.\ the 
dispersion at short times, which arises from the walks along the 
filament, is smaller in discrete than in continuous time. For the 
large time result, which we will derive next, the choice of continuous 
or discrete time makes no difference. The relative 
dispersion 
\begin{equation} 
  \frac{ \Delta n^2_\bd}{\overline{n}^2_\bd}\approx 
  \frac{2}{15}\,\frac{\epsilon\,\sqrt t}{\sqrt\pi} 
\end{equation} 
is small since $t\ll 1/\epsilon^2$.  Using the dispersion $\Delta 
n^2_\bd$, we may also calculate the time-dependent diffusion 
coefficient 
\begin{equation}   
  D_\bd(t)\equiv\half\frac{\d \Delta n^2_\bd}{\d t}\approx 
  \half(1-\gamma)+ 
  \frac{1}{6\sqrt\pi}\,\frac{\va ^2}{\epsilon^2}\,\epsilon^3t^{3/2} 
  +\frac{2\gamma-1}{4}\,\frac{\epsilon\sqrt{t}}{\sqrt{\pi}}. 
\end{equation} 
In the scaling regime $t\sim 1/\epsilon^2$, $D_\bd(t)$ is much larger 
than its limiting value $\half(1-\gamma)$ for $\epsilon=0$.  This 
enhanced diffusion arises from the fact that each motor may detach 
 from the microtubule with probability $\exp(-\tau)$ for any value of 
$\tau=\half\eps\,t$, according to Eq.\ (\ref{punb=}). This leads to a 
considerable broadening of the bound motor distribution. 
 
For large $t$ one can make a change of variables in the integral 
expression (\ref{2ndMomentBound_LF}) and use $y\equiv \eps^2Tx$ as the 
new integration variable. This leads to the asymptotic behavior 
\begin{equation}  
  N_2(t)\approx\frac{4\va ^2\sqrt t}{\epsilon^3\sqrt\pi} 
  (1-\frac{3}{2}\,\frac{\sqrt\pi}{\epsilon\sqrt{t}})+ 
  \frac{1-\gamma}{\epsilon^2}(1-\frac{2}{\epsilon\sqrt{\pi t}})  
\end{equation} 
and 
\begin{equation}  
  \Delta n^2_\bd \approx\frac{\va ^2}{\epsilon^2} 
  (4-\pi-\frac{2\sqrt\pi}{\epsilon\sqrt t})t 
  +\frac{1-\gamma}{\epsilon}\sqrt{\pi t}. 
\end{equation} 
The diffusion coefficient behaves as 
\begin{equation}\label{Dparline}  
  D_\bd(t)\approx 
  \frac{\va ^2}{2\epsilon^2} (4-\pi-\frac{\sqrt\pi}{\epsilon\sqrt t}) 
  +\frac{(1-\gamma)\sqrt{\pi}}{4\epsilon\sqrt{ t}} 
\end{equation} 
for large $t$. 
The limiting value of the diffusion coefficient, 
$D_\bd(\infty)\sim (v_\bd/\eps)^2$ is large compared to the diffusion 
coefficient of the one-dimensional random walk along the filament, 
$D_\bd(0)=(1-\gamma)/2$. This broadening of the distribution occurs 
since the unbound motors lag behind the bound ones, which implies that 
also the rebinding motors lag behind those that have been bound for 
some time.

\subsubsection{The density profile on the filament}

\begin{figure}[h] 
  \begin{center} 
    \leavevmode 
    \includegraphics[angle=-90,width=.5\textwidth]{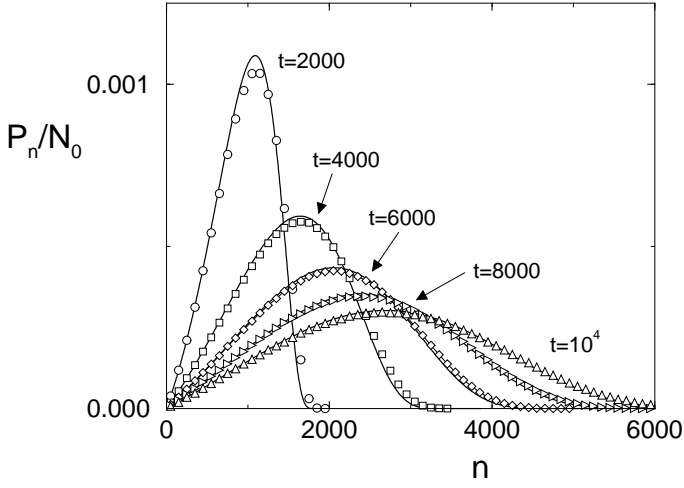}
    \caption{Density profile $P_\bd(n,t)=P_{n,0}(t)$ of motors bound to  
      the filament as a function of the spatial coordinate $n$ 
      parallel to the filament for different times $t$. The profiles 
      are normalized with respect to the probability $N_0$ to be 
      bound at the filament. The lines indicate the analytical result 
      from Eq.\ (\ref{pn0t=}), data points are from Monte Carlo 
      simulations. The parameters are $\epsilon=0.05$ and 
      $\gamma=\delta=0$.} 
    \label{fig:bound_profile_d2} 
  \end{center} 
\end{figure}

At large times $\overline {n^2}_\bd$ scales as $\bar n^2_\bd$ and one 
may look for a scaling form of the density.  For small $s$ and small 
$r$, the Fourier--Laplace transformed probability distribution 
$P_\bd(r,s)$ as in (\ref{sold=2} behaves as 
\begin{equation} \label{p0app} 
  P_\bd(r,s)\approx\frac{1}{s-i\va r+\epsilon\sqrt{s}}. 
\end{equation} 
The inverse Fourier transform follows from the continuum limit of Eq.\  
(\ref{p0rz}). It then follows that, in this limit, $P_{n,0}(s)=0$ for 
$n<0$, while for positive $n$, one finds 
\begin{equation} \label{pn0s} 
  P_{n,0}(s)\approx\frac{1}{\va }e^{-n(s+\epsilon\sqrt{s})/\va } 
\end{equation} 
the inverse Laplace transform of this expression now leads to 
$P_{n,0}(t)=0$ for $n>\va t$ which implies that the overall motion is 
slower than ballistic.  For $n<\va t$, we obtain 
\begin{equation}  
  P_{n,0}(t)\approx\frac{1}{\pi \va }\int_0^\infty \d s\,e^{-s(t-n/\va )} 
  \sin\frac{\epsilon\, n\sqrt{s}}{\va }. 
\end{equation} 
After a change of variables from $s$ to $u=\sqrt{s}$, 
the $u$-integral can be taken over the whole real axis (at the expense 
of a factor $\half$). Evaluation of this Gaussian integral leads to 
\begin{equation} \label{pn0t=} 
  P_{n,0}(t)\approx\frac{\epsilon n}{2\sqrt{\pi \va }(\va t-n)^{3/2}}\, 
  \exp\left(-\frac{\epsilon^2 n^2}{4\va (\va t-n)} \right)
\end{equation} 
for $n\ge 0$. 
This expression vanishes linearly as $\sim n$ for small $n$ and even 
exponentially fast as $n$ approaches $v_\bd t$ from below.  The 
density profile $P_{n,0}(t)$ as given by (\ref{pn0t=}) is plotted in 
Fig.\ \ref{fig:bound_profile_d2} for several values of the time $t$. 
Comparison with the results of MC simulations shows that the 
asymptotic expression as given by (\ref{pn0t=}) is very good for times 
which excess about 8000 time steps. At smaller times, the asymptotic 
expression overestimates the maximum of $P_{n,0}$ and underestimates 
the tails of $P_{n,0}$ for large $n$, see Fig.\  
\ref{fig:bound_profile_d2}. Simulation data are obtained by averaging 
over $5\times 10^7$ realizations of the random walk.

It is somewhat tedious to show that the moments $N_0$, $N_1$ and 
$N_2$ as obtained from (\ref{pn0t=}) agree with the previously derived 
expressions.  To verify this, one may, in an intermediate step, use 
the substitution $n=2\va t(\sqrt{y+y^2}-y)$ which implies 
\begin{eqnarray}  
  P_{n,0}(t)\,\d n & = &\frac{\epsilon\sqrt{t}\,\d y}{\sqrt{\pi(1+y)}} 
  e^{-\epsilon^2 ty}\nn
   & = & \frac{\epsilon^2t\,\d y}{\pi} 
  \int_0^\infty\frac{\d x}{\sqrt{x}}\,e^{-(x+y+xy)\epsilon^2 t}. 
\end{eqnarray}

\subsection{Properties of the unbound motors} 
 
Eventually every motor will unbind and diffuse in the surrounding 
fluid. We now discuss the effects of the filament on the behavior of 
the unbound motors. 
The Fourier--Laplace transform $P(q,r,s)$ of the probability distribution $P_{n,m}(t)$ of bound and unbound motors as given by Eq.\ (\ref{p2=}) can be rewritten as  
\begin{equation}\label{p2decompose}  
  P(q,r,s)=P_\bd(r,s)+P_\ub(q,r,s). 
\end{equation} 
The first part is $P_\bd(r,s)$ as given by Eq.\ (\ref{p0app}), 
describing the bound motors.  The second part describes the 
corresponding probability distribution 
\begin{widetext}
\begin{eqnarray}\label{P_ub_2d=} 
  \lefteqn{ P_\ub(q,r,s) = \sum_{m\neq 0}\sum_{n} e^{iqm+irn}P_{n,m}(s)= \frac{1}{s+1-\half \cos q-\half \cos r}} \nn 
   & & {}+\left[\frac{\gamma(1-\cos r)+\frac{1-\eps}{2}(\cos r-\cos q)+iv_\bd \sin r}{s+1-\half \cos q-\half \cos r}-1\right]P_\bd(r,s). 
\end{eqnarray} 
of the unbound motors.   
For small $s$, small $r$, and small $q$, taking into account that $r\sim s$ and $q\sim\sqrt{s}$, the expression (\ref{p2decompose}) leads to 
\begin{equation}\label{p2ass=}  
  P(q,r,s)=P_\bd(r,s)+P_\ub(q,r,s)\approx\frac{1}{s-i\va r+\epsilon\sqrt{s}} 
  +\frac{4\epsilon\sqrt{s}}{(s-i\va r+\epsilon\sqrt{s})(q^2+4s)}. 
\end{equation} 
\end{widetext}

\subsubsection{Position and longitudinal diffusion} 
 
Expanding (\ref{P_ub_2d=}) or, for large times, (\ref{p2ass=}) in 
powers of $r$ and $q$ yields the moments of the distribution of 
unbound motors. The total fraction of unbound motors as obtained from 
(\ref{p2decompose}) for $r=q=0$ is, of course, $N_0^\ub(t)=1-N_0(t)$ 
expressing motor conservation.  The average longitudinal position of 
the unbound motors 
is given by 
\begin{eqnarray}  
  N_1^\ub(s) & \equiv & \sum_{n}\sum_{m\neq 0}n\,P_{n,m}(s)=-i\frac{\partial}{\partial r}P_\ub(q,r,s)\Big|_{r=q=0}\nn  
   & = & \frac{v_\bd N_0(s)}{s}-N_1(s)\approx\frac{\eps \va }{\sqrt{s}(s+\epsilon \sqrt{s})^2}. 
  \label{N_1ub_s_d=2}  
\end{eqnarray} 
as follows from (\ref{P_ub_2d=}) and (\ref{p2ass=}).  The last 
equality holds for small $\eps$ and small $s$, i.e.\ for large $t$. 
 
For small $t$, one has from the exact expression in 
(\ref{N_1ub_s_d=2}) 
\begin{equation}  
  N_1^\ub(t)\approx\frac{4\eps \va t^{3/2}}{3\sqrt{\pi}}(1-\frac{3}{4}\,\epsilon 
  \sqrt{\pi t}), 
\end{equation} 
which leads to 
\begin{eqnarray}  
  \bar n_\ub & \approx &\frac{2}{3}\va t(1-\oq \epsilon\sqrt{\pi t})\nn
  {\rm and} 
  \qquad \bar v_\ub & \approx & \frac{2}{3}\va (1-\frac{3}{8}\, \epsilon\sqrt{\pi t})
\end{eqnarray} 
for the average position and speed of the unbound motors.  For large 
$t$, the asymptotic expression given in (\ref{N_1ub_s_d=2}) leads to 
\begin{equation}  
  N_1^\ub (t)\approx\frac{\va }{\pi\epsilon^2} 
  \int_0^\infty \d s\,\frac{1-e^{-\epsilon^2t\,s}}{s^{3/2}(s+1)}, 
\end{equation} 
which implies 
\begin{eqnarray}
  N_1^\ub(t) & \approx & \frac{2\va \,\sqrt{t}}{\sqrt{\pi}\epsilon}\, 
  (1-\half\,\frac{\sqrt{\pi}}{\epsilon\sqrt{t}}),\nn 
  \bar n_\ub & \approx & \frac{2\va \,\sqrt{t}}{\sqrt{\pi}\epsilon}, 
  \nn
  {\rm and} \qquad  
  \bar v_\ub(t) & \approx & \frac{\va }{\sqrt{\pi}\epsilon\,\sqrt{t}}. \label{large_t_displ} 
\end{eqnarray} 
Whereas each individual motor has zero average velocity in the fluid, 
the statistical velocity $\bar v_\ub$ is non-zero, since it is driven 
by unbinding from the cloud of motors moving on the filament: The 
cloud of unbound motors advances, because motors rebind to the 
filament and others detach from it, and those detaching have 
propagated a certain distance compared to those rebinding. 
 
Since for large times all motors are detached from the filament most 
of the time, equation (\ref{large_t_displ}) gives the asymptotic 
displacement of a motor, if averages are taken over all bound and 
unbound motors, i.e., 
\begin{equation} 
  \bar v(t)=\bar v_\bd N_0+\bar v_\ub N_0^{\ub}\approx v_\bd N_0^{\ub}=\frac{v_\bd}{\sqrt{\pi}\epsilon \sqrt{t}} 
\end{equation} 
at large times as follows from (\ref{N0t>=}), (\ref{venhd=2}), 
(\ref{large_t_displ}), and $N_0^{\ub}\approx 1$. The effective 
(time-dependent) velocity at large times is $\bar v(t)\approx\va 
N_0(t)$, as predicted by scaling arguments 
\cite{Ajdari1995,Lipowsky__Nieuwenhuizen2001}.  This can also be 
obtained by inspection of Eq.\ (\ref{N_1ub_s_d=2}): The displacement 
of all motors is given by 
\begin{equation} 
  N_1(s)+N_1^\ub(s)=\va\frac{N_0(s)}{s}. 
\end{equation} 
The normalization factor for all motors is unity.  The effective 
velocity is obtained from the inverse: 
\begin{equation} 
  \bar v(t)=\va\frac{\d}{\d t}\int \frac{\d s}{2\pi i}\frac{e^{st}N_0(s)}{s} 
=\va\int \frac{\d s}{2\pi i}e^{st} N_0(s)= \va N_0(t). 
\end{equation} 
 
\begin{figure}[h] 
  \begin{center} 
    \leavevmode 
    \includegraphics[angle=-90,width=.5\textwidth]{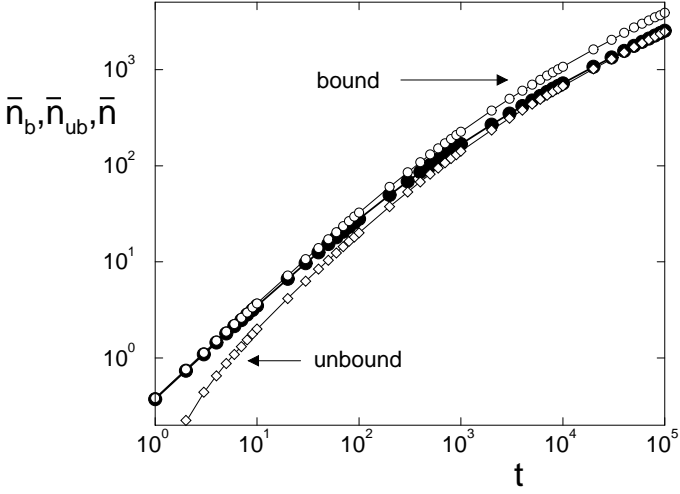}
    \caption{Displacement of motors as a function of time $t$ in the  
      two-dimensional case, as obtained from the exact integrals 
      (lines) and Monte Carlo simulations (data points): Open circles show 
      the average position $\bar n_\bd$ of the bound motors, diamonds 
      the average position $\bar n_\ub$ of the unbound motors, and 
      full circles indicate the displacement $\bar n$ averaged over all 
      motors, which interpolates between the curve for the bound 
      motors at small times and the curve for the unbound motors at 
      large times.  The parameters are $\epsilon=0.05$, $\gamma=0$, 
      and $\delta=0.6$.} 
    \label{fig:displ_d2_b_ub_all} 
  \end{center} 
\end{figure}

\begin{figure}[h] 
  \begin{center} 
    \leavevmode 
    \includegraphics[angle=-90,width=.5\textwidth]{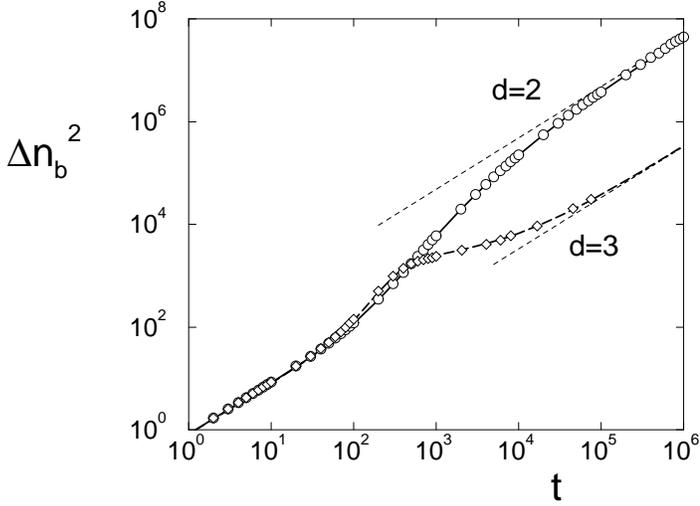}
    \caption{Diffusion of motors parallel to the filament in $d=2$ and $d=3$:  
      Dispersion of bound motors parallel to the filament, $\Delta 
      n_\bd^2$, as a function of time $t$. The dotted lines indicate 
      the linear behavior described by the large-time diffusion 
      coefficient $D_\bd(t=\infty)$ as given by Eqs.\ (\ref{Dparline}) 
      and (\ref{largeTimeDiff_d=3}). In two dimensions, this diffusion 
      coefficient is anomalously high; in three dimensions, it is given 
      by the diffusion away from the filament but exhibits large 
      logarithmic corrections.  The parameters are the same as in 
      Fig.\ \ref{fig:displ_d2_b_ub_all}. } 
    \label{fig:deltanb2_d23} 
  \end{center} 
\end{figure}

Fig.\ \ref{fig:displ_d2_b_ub_all} shows the displacement of the bound 
and unbound motors, obtained by numerical evaluation of the exact 
integrals (dotted lines) and corresponding simulation data (circles 
for the bound motors and diamonds for the unbound ones).  The 
displacement averaged over all motors, $\bar n= \bar n_\bd N_0+\bar 
n_\ub N_0^{\ub}$, is also shown in Fig.\ \ref{fig:displ_d2_b_ub_all} 
(solid line and squares).  For small times it is equal to the 
displacement of the bound motors, for intermediate times it 
interpolates to the curve for the unbound motors. Since almost all 
motors are detached for large times, the displacement of all motors is 
then equal to the displacement of the unbound ones.

The longitudinal position has the second moment 
\begin{eqnarray}  
  N_2^\ub(s) & = & \frac{1}{2s^2}-N_2(s)+\frac{2\va}{s}N_1(s)- 
  \frac{\gamma-\frac{1-\eps}{2}}{s} N_0(s) \nonumber\\ 
  &\approx & \frac{2\eps \va ^2}{\sqrt{s}(s+\epsilon \sqrt{s})^3}. 
\end{eqnarray} 
At short times, this means 
\begin{equation}  
  N_2^\ub(t)\approx\frac{16}{15\sqrt{\pi}}\va ^2\eps t^{5/2}(1-\frac{15}{16}\, 
  \frac{\epsilon\sqrt{t}}{\sqrt \pi}) 
\end{equation} 
and thus 
\begin{eqnarray}  
  \overline{n^2}_\ub & \approx &\frac{8}{15}\va ^2 t^2(1-\frac{7}{16}\, 
  \frac{\epsilon\sqrt{t}}{\sqrt \pi}) \nn
  {\rm and}\qquad 
  \Delta n^2(t) & \approx &\va ^2t^2\frac{4}{45}\,(1-\frac{1}{8}\, 
  \frac{\epsilon\sqrt{t}}{\sqrt{\pi}}). 
\end{eqnarray} 
The last relation implies that, in the longitudinal direction, the 
released motors spread in a broad cloud, not narrowly centered around 
its average. 
  
At long times, one obtains 
\begin{equation}  
  N_1^\ub\approx\frac{2\va \sqrt{t}}{\epsilon\sqrt{\pi}}\qquad{\rm and}\qquad 
  N_2^\ub\approx\frac{2\va ^2t}{\eps^2}. 
\end{equation} 
This implies the dispersion and the diffusion coefficient as given by 
\begin{equation}  
  \label{Dparall}
\Delta n^2_\ub\approx\frac{2(\pi-2)\va ^2t}{\pi\eps^2}\qquad  
  {\rm and} \qquad  
  D_\parallel\approx\frac{(\pi-2)\va ^2}{\pi\eps^2}. 
\end{equation} 
The order of magnitude of the diffusion coefficient is again $\va 
^2/\eps^2$. 
The prefactor
$1-2/\pi\simeq 0.36$ is slightly smaller than the 
prefactor 
$2-\pi/2\simeq 0.43$ of Eq.\ (\ref{Dparline}).  The order of magnitude 
$D_\parallel\sim \va ^2/\eps^2$ tells us that, like on the line, 
longitudinal diffusion is strongly enhanced by the unbinding from and 
binding back to the line.

\subsubsection{Transverse diffusion} 
 
\begin{figure}[h] 
  \begin{center} 
    \leavevmode 
    \includegraphics[angle=-90,width=.5\textwidth]{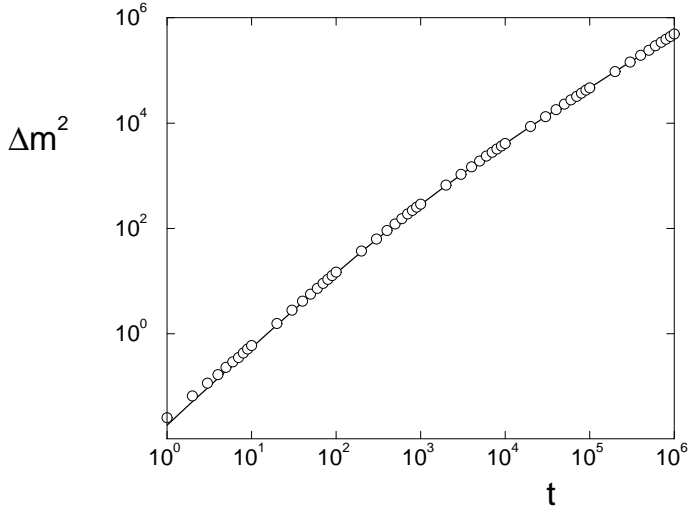}
    \caption{Diffusion of motors perpendicular to the filament in $d=2$:  
      Transverse dispersion $\Delta m^2$ as a function of time $t$. 
      The parameters are the same as in Fig.\  
      \ref{fig:displ_d2_b_ub_all}. } 
    \label{fig:m^2_d2} 
  \end{center} 
\end{figure}

The diffusion behavior perpendicular to the filament is determined by (\ref{p2ass=}) up to quadratic order in $q$.  The average transverse position vanishes and the dispersion of the  
transverse position is given by 
\begin{equation}  
  \Delta m^2(s)=\overline{m^2}(s)=\frac{1}{2s^2}-\frac{1-\eps}{2s}N_0(s)\approx\frac{\eps}{2s^{3/2}(s+\epsilon\sqrt{s})}  
\end{equation} 
and 
\begin{equation}  
  \Delta m^2(t)=\overline{m^2}(t)\approx\frac{\epsilon}{2\pi} 
  \int_0^\infty \d s\,\frac{e^{-st}-1+st}{s^{3/2}(s+\epsilon^2)}. 
\end{equation} 
This implies  
\begin{equation}  
  \Delta m^2(t)\approx\frac{2\epsilon\,t^{3/2}}{3\sqrt\pi}\qquad {\rm and}\qquad 
  D_\perp(t)\approx\frac{\epsilon\,\sqrt{t}}{2\sqrt\pi} 
\end{equation} 
at short times.  
It is small because the motors started in a state bound to the line.  
At large times one has 
\begin{equation}  
  \Delta m^2(t)\approx\half t\,(1-2\frac{1}{\epsilon\sqrt{\pi t}}). 
\end{equation} 
The transverse diffusion coefficient 
\begin{equation}  
  D_\perp(t)\approx\frac{1}{4}(1 - \frac{1}{\epsilon\sqrt{\pi t}}) 
=\frac{1}{4}[1 - N_0(t)] 
\end{equation} 
approaches the free value $\frac{1}{4}$. So the transverse diffusion  
starts out very small, and finally reaches its value in free space. 
The quantity $\Delta m^2(t)$ is plotted in Fig.\ \ref{fig:m^2_d2}. 
There are small deviations in comparison to the simulations at small
 times arising from the time discretization.

\subsubsection{Spatio-temporal density profile of the unbound motors} 
 
Finally, we derive the density profile of the unbound motors.  
After inversion of the Fourier--Laplace transforms, the profile  
(\ref{p2ass=}) becomes in real space 
\begin{equation}  
  P_{n,m}(s)\approx\frac{\eps}{\va }\, 
  \exp\left(-\frac{ns}{\va }-(\frac{n\eps}{\va }+2|m|)\sqrt{s}\right). 
\end{equation} 
Comparing with (\ref{pn0s}) we have a shifted value for $\eps n$ 
and an extra factor $\eps$. The temporal form thus becomes  
\begin{equation} \label{pnmt=} 
  P_{n,m}(t)\approx\frac{\eps(\eps n+2|m|\va )} 
  {2\sqrt{\pi \va }(\va t-n)^{3/2}}\, 
  \exp\left(-\frac{(\eps n+2|m|\va )^2}{4\va (\va t-n)} \right)
\end{equation} 
for $m\neq 0$ and $n\ge 0$. 
This again vanishes for $n\to \va t$. For $n\to 0$ it is finite 
when $m$ is finite, while it is zero for negative $n$, in the present 
approximation. Though detachment followed by 
diffusion to negative $n$ may occur, the attainable values are of 
order unity, and vanish to leading order in $\va /\eps^2$.

\subsection{Pinning line with several tracks} 
 
Microtubules, the filament tracks of kinesin and dynein motors, consist  
of 13 parallel protofilaments \cite{Howard2001}, each of which provides a possible track for these 
motors. To incorporate this in our model, we assume 
that the pinning line at $m=0$ has $k$ internal states where 
$k$ may be equal to $13$. The average occupation of each  
of these states is denoted by $p_{n,0}^{\,j}$ with $j=0,..,k-1$. 
There is a small probability, $\half\zeta$, that a motor 
goes from track $j$ to $j+1$ within one time step,  
and similarly for going to track $j-1$. 
To take into account the cylindrical structure of the microtubule, 
we identify $j=k$ with $j=0$. We assume that after detaching,  
the motor may randomly go to the right or to the left of the tubule; 
likewise, when attaching to the tubule either from right or left, 
the motor has equal probability to attach to any of the tracks. 
 
In this model  the total fraction of motors at position $n$  
along the tubule is 
\begin{equation}  
  P_{n,0}=\sum_{j=1}^kP_{n,0}^{\,j}. 
\end{equation} 
The dynamics given by Eqs.\ (\ref{me1}), (\ref{me2}), 
(\ref{me3}), and (\ref{me4}) remains valid and leads to the same solution, 
as it is insensitive to the internal distribution over the tracks. 
 
The motion on the individual tracks is described by a master equation 
analogous to Eq.\ (\ref{me2}) and given by  
\begin{widetext}
\begin{eqnarray} \label{me2j} 
  P_{n,0}^{\,j}(t+1) & = & \frac{1}{4k} (P_{n,1}+ P_{n,-1}) 
  +(1-\gamma-\half\delta-\half\eps) P_{n-1,0}^{\,j} 
  +\frac{\delta}{2} P_{n+1,0}^{\,j}\nn 
  & & {}  +\frac{\zeta}{2}(P_{n,0}^{\,j+1}+P_{n,0}^{\,j-1}) 
  +(\gamma-{\zeta}) P_{n,0}^{\,j}. 
\end{eqnarray} 
When summed over $j$ this indeed leads back to Eq.\ (\ref{me2}). 
To solve Eq.\ (\ref{me2j}), another Fourier transform is needed which is defined by 
\begin{equation}  
   P_{n,0}^\omega\equiv\sum_{j=1}^k P_{n,0}^{\,j}\,e^{ij\omega} 
\end{equation} 
with $\omega=2\pi \ell/k$ and $\ell=0,..,k-1$. 
We assume that, at $t=0$, all motors are located on the track $j=0$ which corresponds to the initial condition 
\begin{equation}  
  P_{n,m}^{\,j}(t=0)=\delta_{n,0}\delta_{m,0}\delta_{j,0}. 
\end{equation} 
Going to the Fourier--Laplace transform, we find 
\begin{equation}  
  P_0^\omega(r,z)=\frac{2}{2+2s-(2-2\gamma-\delta-\epsilon)e^{ir} 
    -2\gamma-\delta e^{-ir}+2\zeta(1-\cos\omega)}. \label{sold=2om} 
\end{equation} 
The surviving fractions on the individual tracks are 
 found from the expressions for $r=0$. In the time representation, the surviving fractions behave as 
\begin{equation}  
  P^{\,j}(t)=\sum_n P_{n,0}^{\,j}(t)=\frac{1}{k}N_0(t) 
  +\frac{1}{k}\sum_{\omega\neq 0}e^{-ij\omega}\, 
  e^{-\half\epsilon t -\zeta(1-\cos\omega)t}. 
\end{equation}
\end{widetext} 
The first term describes the symmetric distribution of the motors over 
the $k$ tracks; in Eq.\ (\ref{N0t=}) we showed that it decays algebraically.  
The second term represents the asymmetry arising from the initial condition that only one track is populated.  
There are two decay mechanisms of this asymmetry.  
The term $\exp(-\half\epsilon t)$ expresses that detachment and  
subsequent reattachment to randomly chosen tracks restores the symmetry. 
This happens in particular for motors that quickly return to a  
randomly chosen track, implying ordinary exponential decay. 
The factor 
$\exp(-\zeta(1-\cos\omega)t)$ expresses that hopping to neighboring  
tracks also restores the symmetry. For large $k$, the smallest of these 
rates is $2\pi^2\zeta/k^2$. So the exchange process dominates when this  
value is larger than $\epsilon/2$. 
 
Therefore, the asymmetry between the average occupation 
of the various tracks disappears after the discussed transient time.


\section{Random walks in three dimensions} 
\label{sec:3d} 
 
\begin{figure}[h] 
  \begin{center} 
    \leavevmode
    \includegraphics[angle=0,width=.5\textwidth]{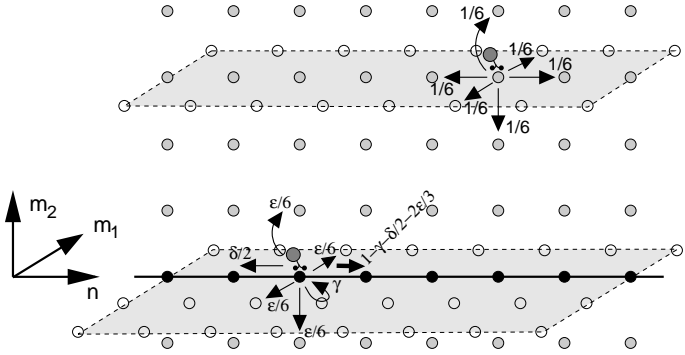}
    \caption{Random walks probabilities for movements in three dimensions: 
      At a non-filament site, a motor jumps with probability $1/6$ to each  
      of the six neighbor sites,  
      at a filament sites, a forward step has probability  
      $1-\gamma-\delta/2-2\epsilon/3$, a backward step 
      $\delta/2$, and a jump to each of the six adjacent non-filament  
      sites has probability $\epsilon/6$; as in two dimensions, a motor  
      does not step at all with probability $\gamma$. 
      As in Fig.\ \ref{fig:randWalk_2d}, the line of black lattice sites  
      represents the filament. The shaded ares with white lattice sites  
      indicate the planes spanned by the  $n$- and $m_1$ axes of the lattice  
      perpendicular to the  
      paper plane. } 
    \label{fig:randWalk_3d} 
  \end{center} 
\end{figure}

Now, let us consider the same kind of random walk on a 
three-dimensional cubic lattice, in which the line $m_1=m_2=0$ 
represents the filament which attracts and binds the motors.  Away 
 from the filament the jump probabilities are equal to $\os$, while on 
the filament they are given by $1-\gamma-\half 
\delta-\frac{2}{3}\epsilon$ and $\half \delta$ in the forward and 
backward direction, respectively, equal to $\os\epsilon$ for each of 
the four sideward directions, and equal to $\gamma$ to make no step, 
see Fig.\ \ref{fig:randWalk_3d}. 
The average short-time velocity is $\va 
=1-\gamma-\delta-\frac{2}{3}\eps$ while the sticking probability is 
$\exp[-\frac{2}{3}\eps\,t]$. 
 
We denote the transverse coordinate as $\mt=(m_1,m_2)$.  The master 
equation for this dynamics has the form 
\begin{widetext}
\begin{eqnarray} \label{me31} 
  P_{n,\mt}(t+1)&=&\os P_{n+1,\mt}+\os P_{n-1,\mt}+ 
  \os \sum_\rt P_{n,\mt+\rt} 
  \qquad\qquad (\mt\neq \nulb,\rt)\\  
  \label{me32} P_{n,\nulb}(t+1)&=&\os\sum_\rt P_{n,\rt}+ 
  (1-\gamma-\frac{2}{3}\epsilon-\half\delta)P_{n-1,\nulb} 
  +\half\delta\, P_{n+1,\nulb}+\gamma\,P_{n,\nulb}\\  
  \label{me33} P_{n,\rt}(t+1)&=&\os P_{n+1,\rt}+\os P_{n-1,\rt} 
  +\os\sum_{\rt'(\neq\rt)} P_{n,\rt-\rt'}+\frac{1}{6}\epsilon\,P_{n,\nulb}. 
\end{eqnarray} 
\end{widetext}
In these equations, $\mathbf{\rho}$ and $\mathbf{\rho}'$ denote the 
four transverse nearest neighbor vectors which connect a filament site 
to the four adjacent non-filament sites, $\rt=(0,\pm1),\,(\pm 1,0)$. 
The summations over $\rho$ in (\ref{me31}) and (\ref{me32}) or $\rho'$ 
run over the four possible values. Eq.\ (\ref{me33}) holds for any of 
the four values of $\rho$ with the sum over $\rho'$ running over the 
other three vectors.  We can follow the same steps as in the 
two-dimensional case using again Fourier--Laplace transforms.  The 
Fourier transformation in the perpendicular directions leads to a 
transverse Fourier vector $\qb=(q_1,q_2)$.  The equivalent of Eq.\  
(\ref{p2=}) becomes 
\begin{widetext}
\begin{equation} \label{p3=} 
P(\qb,r,s)=\frac{ 3+[3\gamma+\half(5-6\gamma-4\epsilon-3\delta)e^{ir} 
  -\half(1-3\delta)\,e^{-ir}-(1-\epsilon)(\cos q_1+\cos q_2)]\,  
  P_{\rm b}(r,s)}{3+3s-\cos r-\cos q_1-\cos q_2}. 
\end{equation} 
By doing the integrals over $q_1$ and $q_2$, we derive the expression 
for $P_{\rm b}(r,s)$.  The necessary integral is 
\begin{eqnarray}  
  I(r,s) & = & \int_0^{2\pi}\frac{\d q_1}{2\pi}\,\int_0^{2\pi}\frac{\d q_2}{2\pi}\, \frac{1}{3+3s-\cos r-\cos q_1-\cos q_2}\nn
   & = & \frac{\sqrt{m}}{\pi}\,K(m)\nonumber\\ 
  {}& & {\rm with} \qquad m\equiv\frac{4}{(3+3s-\cos r)^2},\label{integral_d3} 
\end{eqnarray} 
where $K(m)=\int_0^{\pi/2}\d\phi/\sqrt{1-m\sin^2\phi}$ is the complete 
elliptic integral of the first kind.  We also use the relation 
\begin{equation} 
  \int_0^{2\pi}\frac{\d q_1}{2\pi}\,\int_0^{2\pi}\frac{\d q_2}{2\pi}\, 
  \frac{\cos q_1+\cos q_2}{3+3s-\cos r-\cos q_1-\cos q_2} 
  =(3+3s-\cos r)I(r,s)-1, 
\end{equation} 
where $I(r,s)$ is given by (\ref{integral_d3}).  We then get 
\begin{equation} \label{pnulb_d3} 
  P_{\rm b}(r,s)=\frac{3\,I(r,s)}{\epsilon+ 
    [3(1-\epsilon)s+ 
    \half(\epsilon-3\delta)\,(e^{-ir}-1) 
    -\half(6-6\gamma-3\delta-5\epsilon)(e^{ir}-1)] 
    \,I(r,s)}. 
\end{equation}
\end{widetext}
For large $s$, one may use $I(s)\approx 1/(3s)$, to verify that 
$P_{\rm b}(r,s)\approx 1/s$, which is required by our initial 
condition that all motors started at $m_1=m_2=n=0$.

\subsection{Behavior on the filament} 
 
For small $\epsilon$ and small $r$ and $s$, we may approximate 
(\ref{pnulb_d3}) as 
\begin{equation}\label{p03d=}  
  P_{\rm b}(r,s)\approx\frac{1}{s-i\va r+J(s)}\qquad {\rm with}\qquad 
  J(s)\equiv\frac{\epsilon}{3I(0,s)}. 
\end{equation}  
which is analogous to (\ref{p0app}).  This implies in real space that 
\begin{equation}  
  P_{n,\nulb}(s)\approx\frac{1}{\va }\,e^{-n(s+J(s))/\va } 
\end{equation} 
for $n\ge 0$.  Decomposing $J(-s\pm i0)=J_1(s)\pm iJ_2(s)$, the 
inverse Laplace transform leads to 
\begin{equation}  
  P_{n,\nulb}(t)\approx\int_0^\infty\frac{\d s}{\pi \va } 
  \,e^{-s(t-n/\va )-nJ_1(s)/\va } \sin \frac{nJ_2(s)}{\va }. 
\end{equation}

For large times, the variable $s$ will be small. 
Since $K(m)\approx (1/2)\ln[16/(1-m)]$ for $m\approx 1$ 
\cite{Abramowitz}, we may conclude that $I$ diverges as 
\begin{equation}\label{int+def_taun} 
  I(r,s)\approx-\frac{1}{2\pi}\ln[\taun(s+\os r^2)]\qquad {\rm with}\qquad \taun=\frac{3}{16} 
\end{equation} 
for small $s$ and $r$. This implies in particular that 
\begin{eqnarray}  
  J(s) & \approx & -\frac{2\pi\epsilon}{3\ln \taun s},\nn 
  J_1(s) &\approx &-\frac{2\pi\epsilon}{3}\, \frac{ \ln\taun s}{[\ln\taun 
    s]^2+\pi^2},\nn
  {\rm and}\qquad J_2(s) & \approx & \frac{2\pi\epsilon}{3}\, 
  \frac{\pi}{[\ln\taun s]^2+\pi^2}.  \label{J12ass}
\end{eqnarray}

\begin{figure}[h] 
  \begin{center} 
    \leavevmode 
    \includegraphics[angle=-90,width=.5\textwidth]{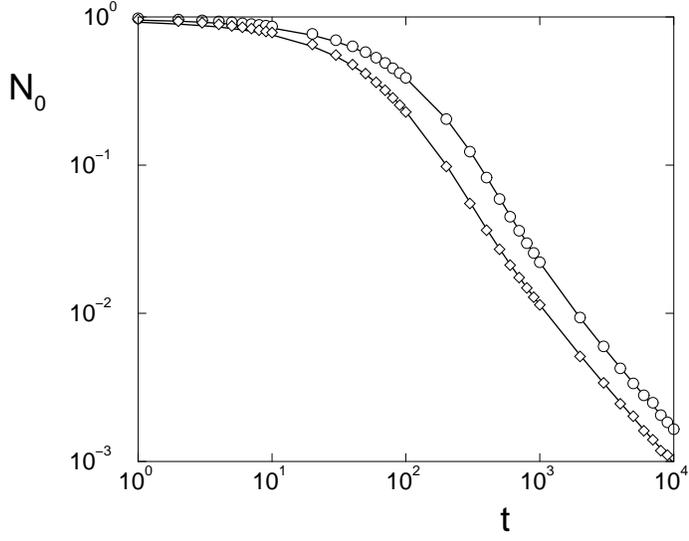}
    \caption{Fraction of bound motors, $N_0$, as a function of time for the  
      three-dimensional case. The two curves are for $\epsilon=0.03$ 
      (circles) and $\epsilon=0.05$ (diamonds). The other parameters 
      are $\gamma=0$ and $\delta=0.6$.} 
    \label{fig:Pb_d3_eps.03.05} 
  \end{center} 
\end{figure}

The Laplace transformed survival fraction $N_0(s)$ is equal to 
$P_\bd(0,s)$. It then follows from (\ref{p03d=}) that 
\begin{equation} \label{N0_d=3_exact}
  N_0(t)=\int_0^{\infty}\frac{\d s}{\pi}\, 
  \frac{e^{-st}J_2(s)}{[-s+J_1(s)]^2+J^2_2(s)}.  
\end{equation}  
For $t\gg 1/\epsilon$, using (\ref{J12ass}) and neglecting $s$ with 
respect to $\ln s$, we get for the number of particles on the line the 
simple result 
\begin{equation} N_0(t)\approx 
  \int_0^\infty\frac{\d s}{\pi}\,\frac{3\,e^{-st}}{2\epsilon}= 
  \frac{3}{2\pi\epsilon t}, 
\end{equation} 
which confirms the scaling $N_0(t)\sim t^{-1}$ predicted by the 
scaling approach \cite{Lipowsky__Nieuwenhuizen2001}.  We can also 
derive the latter result from $N_0(s)=[{3}/({2\pi\eps})]\ln(\taun s)$ 
using the Tauberian transforms summarized in appendix 
\ref{sec:app:TaubThms}. The survival fraction is shown in Fig.\  
\ref{fig:Pb_d3_eps.03.05} for two values of $\epsilon$. Again the 
exact integral (\ref{N0_d=3_exact}) is evaluated numerically (lines) 
and compared to simulation data (data points). The agreement is good.

\begin{figure}[h] 
  \begin{center} 
    \leavevmode 
    \includegraphics[angle=-90,width=.5\textwidth]{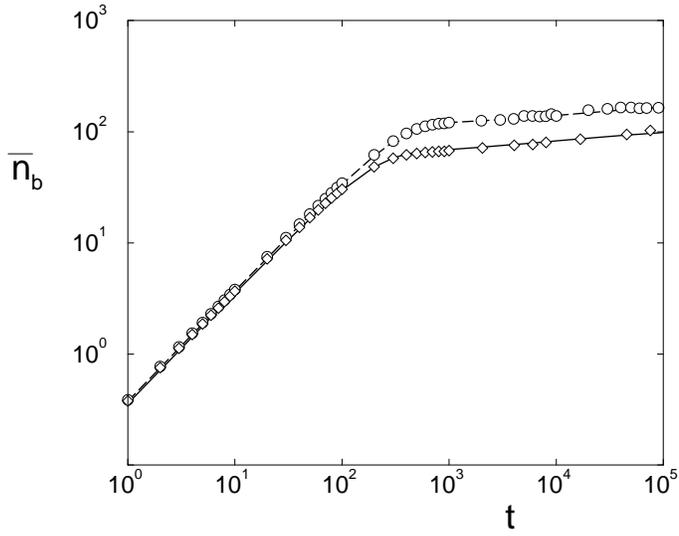}
    \caption{Average position $\bar n_\bd$ of the motors bound to the  
      filament as a function of time $t$ in three dimensions. The  
      parameters are the same as in Fig.\ \ref{fig:Pb_d3_eps.03.05}. } 
    \label{fig:n0_d3_eps.03.05} 
  \end{center} 
\end{figure}

Computation of the first moment $N_1(s)=-i\frac{\partial}{\partial 
  r}P_\bd(r,s)|_{r=0}$ with $P_\bd$ as given by (\ref{p03d=}) leads to 
\begin{equation}  
  N_1(t)=2\va 
  \int_0^{\infty}\frac{\d s}{\pi }\, {e^{-st}}\frac{[J_1(s)-s]J_2(s)} 
  {\left([J_1(s)-s]^2+J_2^2(s)\right)^2}. 
\end{equation}  
For small times, this behaves again as $\va t$ corresponding to the 
velocity $\va=1-\gamma-\delta-\frac{2}{3}\eps$.  For large times, we 
find 
\begin{eqnarray}  
  N_1(s) & \approx &\va 
  \left(\frac{3}{2\pi\eps}\right)^2\ln^2 (s\taun)\nn
  {\rm and} \qquad N_1(t)& \approx & 
  \frac{9\va }{2\pi^2\epsilon^2}\int_0^\infty\d s\,e^{-st}\, \ln 
  \frac{16}{3s} \nn
   & = & \frac{9\va }{2\pi^2\epsilon^2\,t} (\ln \frac{16\, 
    t}{3}+\gamma_E) 
\end{eqnarray} 
to leading order in $1/t$ 
where $\gamma_E\equiv 0.577215$ is Euler's constant.  This implies the 
average position and velocity 
\begin{equation}\label{nb3d=}  
  \bar n_\bd(t)\approx\frac{3\va 
    }{\pi\epsilon}(\ln \frac{t}{\tau_0}+\gamma_E)\quad {\rm and} 
  \quad \bar v_\bd(t)\approx\frac{3\va 
    }{\pi\epsilon\,t}=2N_0(t)\va. 
\end{equation} 
The position of bound motors as a function of time is shown in Fig.\  
\ref{fig:n0_d3_eps.03.05}. The agreement between the analytical result 
and the simulations is again quite good.

For the second moment, we obtain 
\begin{equation} 
  N_2(s)=\frac{2\va^2}{[s+J(s)]^3}+\frac{1-\gamma-\eps}{\va}N_1(s)+\frac{\frac{1}{2\pi\eps 
    s} J^2(s)}{[s+J(s)]^2},  
\end{equation}  
where we have taken into account the quadratic correction term to 
$I(s,r=0)$, and 
\begin{widetext}
\begin{eqnarray}  
  N_2(t) & = & 
  \frac{2\va^2}{\pi}\int_0^\infty\d s\,\frac{e^{-st}[3(-s+J_1(s))^2 
    J_2(s) -J_2^3(s)]}{[(-s+J_1(s))^2+J_2^2(s)]^3}
  + \frac{2(1-\gamma-\eps)}{\pi}\int_0^\infty\d s\,\frac{e^{-st}[(J_1(s)-s)J_2(s)]}{[(-s+J_1(s))^2+J_2^2(s)]^2}\nonumber\\ 
  & & {}- \frac{1}{\pi^2\eps}\int_0^\infty\d 
  s\,\frac{e^{-st}[J_2^3(s)-sJ_1(s)J_2(s)+J_1^2(s)J_2(s)]}{[(-s+J_1(s))^2+J_2^2(s)]^2} 
  + \frac{1}{2\pi\eps} . 
\end{eqnarray}  
\end{widetext}
The last term emerges from the singularity at $s=0$ in the third term 
of $N_2(s)$ and represents diffusion in the unbound state.  For large 
times, this expression leads to the asymptotic relations 
\begin{eqnarray}  
  N_2(s) & \approx &-2\va 
  ^2\left(\frac{3}{2\pi\eps}\right)^3\ln^3 s\taun + \frac{1}{2\pi\eps 
    s},\\
  N_2(t)& \approx & 2\va ^2\left(\frac{3}{2\pi\eps}\right)^3 
  \,\frac{1}{t}\,[3 (\ln 
  \frac{t}{\taun}+\gamma_E)^2-\frac{\pi^2}{3}]+\frac{1}{2\pi\eps}, \\
  \overline{n^2}(t) &\approx &2\va ^2\left(\frac{3}{2\pi\eps}\right)^2 \,[3 
  (\ln \frac{t}{\taun}+\gamma_E)^2-\frac{\pi^2}{3}]+\frac{1}{3}t,
\end{eqnarray} 
and 
\begin{eqnarray} 
  \Delta {n^2}(t) &\approx &2\va ^2\left(\frac{3}{2\pi\eps}\right)^2 \,[(\ln 
  \frac{t}{\taun}+\gamma_E)^2-\frac{\pi^2}{3}]+\frac{1}{3}t.  
\end{eqnarray}  
with $\taun=3/16$ as in (\ref{int+def_taun}).  The longitudinal 
diffusion coefficient is 
\begin{equation} \label{largeTimeDiff_d=3} 
  D_{\rm b}(t)\approx \frac{1}{6}+\frac{9\va^2}{2\pi^2\eps^2\,t}\, \,[\ln 
  \frac{t}{\taun}+\gamma_E]. 
\end{equation}  
Notice that, at typical times $t\sim {1}/{\eps}$, this is still of 
order ${1}/{\eps}$, and, thus, much larger than the value without the 
unbinding mechanism. In contrast to the two-dimensional case discussed 
above, the leading term in three dimensions is given by the usual 
diffusion in the unbound state, but there are large logarithmic 
corrections of order $(\va/\eps)^2$. This can be seen in Fig.\  
\ref{fig:deltanb2_d23}, where $\Delta n_\bd$ is shown for both cases.

\subsubsection{Density profile of the bound motors} 
 
As in the two-dimensional case, the spatial distribution of the motors 
bound to the line can be derived in a somewhat explicit form.  For 
$n\ge 0$ one has 
\begin{equation} \label{profile3d_bound} 
  P_{n,\bf 0}(t)\approx\int_{-i\infty}^{i\infty}\frac{\d s}{2\pi i 
    \va }\,e^{-A}\ \quad {\rm with}\quad  A= {\frac{n}{\va }[s+J(s)]}-st. 
\end{equation}  
For large $n$ and $t$, we may use the saddle point approximation.  The 
condition $A'=0$ yields 
\begin{equation} \label{saddle_point_cond} 
  s=\frac{2\pi\eps n}{3(\va t-n)\ln^2\taun s}. \label{ssp=} 
\end{equation}  
For very small $\eps$ and $\va t-n\sim n$ this means that $s$ is 
indeed small, 
\begin{equation}  
  s\approx\frac{2\pi\eps n}{3(\va 
    t-n)\ln^2\eps}. 
\end{equation}  
The saddle point values are 
\begin{eqnarray} 
  A&=&-\frac{2\pi\eps n}{3\va \ln\taun s}(1+\frac{1}{\ln\taun s}) 
  \approx \frac{2\pi\eps n}{3\va \ln(1/\eps)}\\ 
  -A''&=&\frac{2\pi\eps n}{3s^2\va \ln^2\taun s}(1+\frac{2}{\ln\taun s}) 
  \approx \frac{2\pi\eps n}{3s^2\va \ln^2\eps}.  
\end{eqnarray}  
The second derivative has a negative sign, which allows us to choose 
the contour from $s_{s.p.}-i\infty$ to $s_{s.p.}+i\infty$, where 
$s_{s.p.}$ is the saddle point value given in (\ref{ssp=}).  The 
integration over Gaussian fluctuations yields 
\begin{eqnarray}
  P_{n,\bf 0}(t) & \simeq & \sqrt{\frac{-1}{2\pi A''\va ^2}}\,\,e^{-A}\nn
  & \approx & \frac{\sqrt{\eps n}}{(\va t-n)\sqrt{3\va }}\, 
  \exp\left(-\frac{2\pi\eps}{3\va \ln (1/\eps)}\,n\right). \label{profile_3d}  
\end{eqnarray}  
As a function of $n$, this curve starts at $0$, has a maximum and goes 
to zero at $n=\va t$.  The apparent divergence near $n=\va t$ of the 
last expression is an artifact of the saddle point approximation.

We can check the normalization: 
\begin{equation}  
  \int_0^{\va t/2}\d n\,P_{n,\bf 0}(t). 
\end{equation}  
It differs from the exact value by a factor of order 
$\sqrt{\ln(1/\eps)}$, which is anyhow only logarithmic. This occurs 
because the small $n$-behavior has not been treated properly by the 
saddle point approximation.

\begin{figure}[h] 
  \begin{center} 
    \leavevmode 
    \includegraphics[angle=-90,width=.5\textwidth]{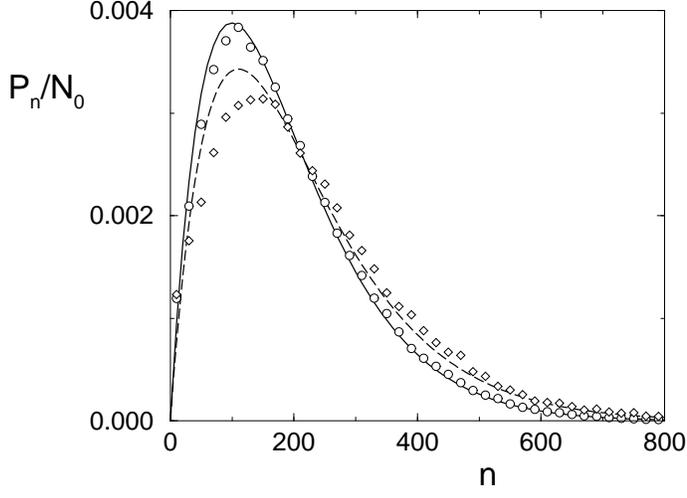}
    \caption{Density profile $P_\bd(n)=P_{n,0}$ of motors bound to the  
      filament for the three-dimensional case as a function of the 
      spatial coordinate $n$ parallel to the filament at times 
      $t=2000$ (solid line, circles) and $t=10^4$ (dashed line, 
      diamonds).  
      Lines are obtained by numerical evaluation of the approximate 
      analytical expression (\ref{profile3d_bound}), data points are 
      from Monte Carlo simulations.  The parameters are the same as in Fig.\  
      \ref{fig:displ_d2_b_ub_all}.  
      } 
    \label{fig:bound_profile_d3} 
  \end{center} 
\end{figure}

We have compared the profiles of the bound motors obtained from the 
saddle point approximation with simulation data, evaluating Eq.\ 
(\ref{profile_3d}) with $s$ taken from the numerical solution of Eq.\ 
(\ref{saddle_point_cond}) and normalizing the saddle point profile. 
As expected, agreement with the simulation data is only good for large 
$n$.  Therefore we have also taken the inverse Laplace transform of 
(\ref{profile3d_bound}) numerically.  The result is shown in Fig.\  
\ref{fig:bound_profile_d3} for $t=2000$ and $t=10000$ in comparison 
with simulation results.  To obtain the simulated profiles, 
simulations were performed again with $5\times 10^7$ motors particles, 
most of which are, however, detached from the filament at the times 
the density profile was measured.  While agreement is good for large 
$n$, there are deviations in the region around the maximum, which are 
probably due to the approximation used in Eq.\ (\ref{p03d=}).

\subsection{Behavior of the unbound motors} 
 
Also for the motors that detached from the pinning line, we can follow 
the steps of previous section. At large times, the transport 
properties obtained for the unbound motors dominate also the results 
that are obtained if averages are taken over all, bound and unbound, 
motors, because at large times the motors spend most of the time 
detached from the filament

\subsubsection{Average position and dispersion of unbound motors}  
 
From Eq.\ (\ref{p3=}), we now find the Fourier--Laplace transform of 
the distribution of unbound motors to be given by 
\begin{equation} \label{pubd=3} 
  P_\ub(\qb,r,s)\approx\frac{J(s)}{(s+\os q^2)[s-ir\va +J(s)]} 
\end{equation} 
for small $r$, $\qb$, and $s$.  The term linear in $r$ implies 
\begin{eqnarray} 
  N_1^\ub(s) & = & \frac{\va}{s}N_0(s)-N_1(s)\approx\frac{\va J(s)}{s\,[s+J(s)]^2}\nn
  {\rm and}\qquad 
  \overline{n}_\ub & \approx & \frac{3\va }{2\pi\eps}(\ln\frac{t}{\taun}+\gamma_E). 
\end{eqnarray} 
The last result is just half of the value for the bound particles 
(\ref{nb3d=}).  As in the two-dimensional case, the latter results 
gives also the average position of all, bound and unbound, motors at 
large times. The logarithmic growth of the average position and the 
corresponding time-dependent velocity confirm the predictions of 
scaling arguments, which give $\bar n(t)\sim \ln t$ 
\cite{Ajdari1995,Lipowsky__Nieuwenhuizen2001}. The time-dependent 
velocity is again given by $v_\bd N_0(t)$ at large times. 
 
For the second moment we get 
\begin{equation} 
  N_2^\ub(s)=\frac{1}{3s^2}-N_2(s)+\frac{2-\gamma-\frac{2\eps}{3}}{s}N_0(s) 
  +\frac{2\va}{s}N_1(s). 
\end{equation} 
For small $s$ or large $t$, this expression leads to the asymptotic 
relations 
\begin{eqnarray}  
  N_2^\ub(s) & \approx &\frac{1}{3s^2}+\frac{2v^2_0J(s)}{s\,[s+J(s)]^2},\nn
  \overline{ n^2}_\ub & \approx &\frac{t}{3}+\frac{9\va ^2}{2\pi^2\eps^2} 
  [(\ln\frac{t}{\taun}+\gamma_E)^2-1], \nn
  \Delta n^2_\ub & \approx & \frac{t}{3}+\frac{9\va ^2}{4\pi^2\eps^2} 
  [(\ln\frac{t}{\taun}+\gamma_E)^2-2],\qquad {\rm and}\nn
  D_\parallel & \approx & \frac{1}{6}+\frac{9\va ^2}{4\pi^2\eps^2\,t} 
  [\ln\frac{t}{\taun}+\gamma_E], 
\end{eqnarray}  
where the logarithmic correction to the free diffusion coefficient 
is half of the correction to the bound diffusion coefficient 
(\ref{largeTimeDiff_d=3} ).  
That such leading singularities have different numerical prefactors
in different quantities could have been anticipated,  
since the cloud of random walkers is smeared, 
with spread as large as the average. In $d=2$ the effect was stronger, 
since already the leading terms had different numerical prefactors,
cf. Eq. (\ref{Dparline}) for   
  $D_\bd$ with Eq. (\ref{Dparall}) for $D_\parallel$.

For transversal transport the situation is much simpler 
\begin{equation}  
  \Delta m_1^2(s)=\Delta m_2^2(s)={\overline m_1^2}(s)={\overline m_2^2} 
  =\frac{J(s)}{3s^2[s+J(s)]}. 
\end{equation} 
For large $t$ one has small $s$, so $s\ll J(s)$. This implies 
\begin{equation}  
  \Delta m_{1,2}^2(t)={\overline m_{1,2}^2}(t)\approx\frac{t}{3}- 
  \frac{1}{2\pi\eps}(\ln\frac{t}{\taun}+\gamma_E) 
\end{equation} 
and the transverse diffusion coefficient is given by 
\begin{equation} 
  D_\perp=\frac{1}{6}-\frac{1}{4\pi\eps t}=\frac{1}{6}[1-N_0(t)]. 
\end{equation} 
The limiting values are just those without the pinning line.  They are 
reached for $t\gg 1/\eps$.

\subsubsection{Density profile of unbound motors} 
 
Finally, we derive an expression for the probability distribution (or, 
equivalently, the density profile in the case of many non-interacting 
motors) away from the filament using again the saddle point 
approximation.  From (\ref{pubd=3}) we deduce 
\begin{equation}  
  \int\frac{\d^2 q}{(2\pi)^2}\,\frac{e^{i\qb\cdot \mb}}{q^2+6s} 
  =\int_0^\infty \frac{\d q\,q}{2\pi}\,\frac{J_0(qm)}{q^2+6s} 
  =\frac{1}{2\pi}\,K_0(m\sqrt{6s}),  
\end{equation} 
for $m=|\mb|\neq 0$, where $K_0$ is a modified Bessel function.  
  This leads to the profile 
\begin{equation}  
  P_{n,\mb}(t)=\frac{3}{\pi \va }\int\frac{\d s}{2\pi i}\,e^{st} 
  K_0(m\sqrt{6s})\,e^{-n[s+J(s)]/\va }. 
\end{equation} 
For closing the contour 
around the negative real axis we need 
\begin{equation}  
  K_0(iz)=\frac{\pi i}{2}H_0^{(1)}(-z)= 
  -\frac{\pi i}{2}H_0^{(2)}(z)=-\frac{\pi }{2}[Y_0(z)+iJ_0(z)]. 
\end{equation} 
We get a result of the type 
\begin{eqnarray}  
  \lefteqn{ P_{n,\mb}(t) = \frac{3}{2\pi \va }\int_0^\infty 
  \d s\,e^{-st+n[s-J_1(s)]/\va } }\nonumber\\
   & & \times
  [\cos \frac{nJ_2(s)}{\va }\,J_0(m\sqrt{6s})- 
  \sin \frac{nJ_2(s)}{\va }\,Y_0(m\sqrt{6s})].
\end{eqnarray} 
The saddle point approximation can be done for large enough $m$, where 
\begin{equation}  
  K_0(z)\approx \sqrt\frac{\pi}{z}\,\,e^{-z}. 
\end{equation} 
We then get for the distribution away from the filament 
($\mb\neq\mathbf{0}$) 
\begin{eqnarray}  
  P_{n,\mb}(t) & = &\frac{3}{\pi \va }\int\frac{\d s} 
  {2i(6s)^{1/4}\sqrt{\pi m}}\,\,e^{-A}\nn
   & \simeq & 
  \frac{3}{\pi \va \,(6s)^{1/4}\sqrt{-2mA''}}\,\,e^{-A}, 
\end{eqnarray} 
where the saddle point values are 
\begin{eqnarray}  
  A&=&\frac{1}{\va }[(n-\va t)s+nJ(s)+m\va \sqrt{6s}] \nn 
  &=&-\frac{2\pi\eps n}{3\va \ln\taun s}(1+\frac{1}{\ln\taun s}) \nn
  & & {}+\frac{2\pi\eps n}{3\ln^2\taun s}\, 
  \frac{m\va \sqrt{3/(2s)}}{\va t-n-\va m\sqrt{3/(2s)}} 
\end{eqnarray} 
and 
\begin{eqnarray}
  -A''&=&\frac{2\pi\eps n}{3s^2\va \ln^2\taun s}(1+\frac{2}{\ln\taun s}) 
  +\frac{m\sqrt{3/8}}{s^{3/2}} \nn
   & \approx & \frac{2\pi\eps n}{3s^2\va \ln^2\eps}+\frac{m\sqrt{3/8}}{s^{3/2}}
\end{eqnarray} 
with the saddle point value of $s$ as given by 
\begin{equation}  
  s=\frac{2\pi\eps n\va }{3\ln^2\tau_0s\,[\va t-n-\va m\sqrt{3/(2s)}]}.  
\end{equation}

\subsection{Pinning line with several tracks} 
 
As before we can assume that the line has $k$ internal states, 
which for $k=13$ model the protofilaments on the microtubule. 
The motion on the individual tracks is described by a master equation 
analogous to Eq.\ (\ref{me2}), 
\begin{widetext}
\begin{eqnarray}
  P_{n,\nulb}^{\,j}(t+1) & = & \frac{1}{6k} \sum_\rho P_{n,\rho} 
  +(1-\gamma-\half \delta-\frac{2}{3}\,\epsilon)\,P_{n-1,\nulb}^{\,j} 
  +\half \delta \, P_{n+1,\nulb}^{\,j} \nn 
  & & {}
  +\frac{\zeta}{2}(P_{n,\nulb}^{\,j+1}+P_{n,\nulb}^{\,j-1})+ 
  (\gamma-\zeta) P_{n,\nulb}^{\,j}.  \label{me3j} 
\end{eqnarray} 
This can be analyzed as in the $d=2$ case. One finds for $\omega\neq 0$ 
\begin{equation}  
  P_{\bf 0}^\omega(r,s)=\frac{1}{s+1-(1-\gamma-\half \delta- 
    \frac{3}{2}\epsilon)e^{ir} 
    -\gamma-\half\delta e^{-ir}+\zeta(1-\cos\omega)} \label{sold=3om}. 
\end{equation} 
\end{widetext}
The surviving fractions on the individual protofilaments are 
 found by inserting $r=0$. In the time representation 
it reads  
\begin{equation}  
  P^{\,j}(t)=\sum_n P_{n,{\bf 0}}^{\,j}(t) 
  =\frac{1}{k}N_{ 0}(t)  
  +\frac{1}{k}\sum_{\omega\neq 0}e^{-ij\omega}\, 
  e^{-\frac{3}{2}\epsilon t -\zeta(1-\cos\omega)t}. 
\end{equation} 
The asymmetry decays 
as $\exp[-(\frac{2}{3}\eps+\zeta-\zeta\cos\omega)\,t]$.  
As in the two-dimensional situation, the asymmetry has no influence  
on the total occupation  
$P_{n,\nulb}=\sum_j\,P_{n,\nulb}^{\,j}$ of site $n$ along the 
microtubule, that was the subject of study in previous subsections.

 
\section{Variable sticking probability} 
\label{sec:sticking_prob} 
 

We now want to incorporate the possibility that a motor need not  
bind to the line when it collides with it. We consider two approaches. 
 
\subsection{Tubule above a two-dimensional plane} 
 
Let us now consider a tubule located above the line $m=0$ of a 
two-dimensional plane.  We consider the following jump rates: 
Detaching: line$\to$plane $\half\eps$; attaching plane$\to$ line 
$\half\eta$; reduced jumps in the plane away from position below the 
line: $(1-\eta)/4$.  The probability to jump from the tubule to the 
line with $m=0$ is equal to $\half\epsilon$ and the probability to 
jump from the line with $m=0$ to the tubule is given by $\half\eta$. 
In addition, the motor particle jumps with probability $(1-\eta)/4$ 
  from the line with $m=0$ to a neighboring line with $m=\pm 1$.  So for 
$\eta=0$ no reattachment occurs. For $\eta=1$ the other extreme 
occurs: one cannot jump from the line $m=0$ to the rest of the plane; 
if initially all walkers were on the tubule, they will go no further 
than below it, but not wander in the plane. 
 
Thus we consider the master equations 
\begin{widetext}
\begin{eqnarray}  
  P_{nm}(t+1)&=&\frac{1}{4}\sum_\rho P_{n+\rho_1,m+\rho_2}(t) 
  +\delta_{m,0}\half\epsilon P_n(t)- 
  \frac{\eta}{4}(\delta_{m,1}+\delta_{m,-1})P_{n,0}(t)\\ 
  P_n(t+1)&=&\gamma P_n(t)+(1-\gamma-\half\delta-\half\epsilon)P_{n-1}(t)+ 
  \half\delta P_{n+1}(t)+\half\eta P_{n,0}(t). 
\end{eqnarray} 
\end{widetext}
As initial condition we choose all motors on the location $n=0$ of the 
tubule, 
\begin{equation}  
  P_{nm}(0)=0\qquad {\rm and} \qquad P_n(0)=\delta_{n,0}. 
\end{equation} 

The Fourier--Laplace transforms yield 
\begin{eqnarray}  
  \lefteqn{[1+s-\half\cos r-\half\cos q]P(q,r) }\nonumber\\
  & & =\half\eps P(r)- \half\eta\cos q P_0(r) 
\end{eqnarray}
\begin{eqnarray}  
   & [1+s-\gamma-(1-\gamma-\half\delta-\half\eps)e^{ir}-\half\delta e^{-ir}]P(r) &  \nonumber\\
  & = 1+\half \eta P_0(r). & 
\end{eqnarray} 
We can integrate $P(q,r)$ over $q$ 
\begin{eqnarray}  
  P_0(r) & = &\int\frac{\d q}{2\pi}\,\frac{\eps P(r)-\eta\cos q P_0(r)} 
  {2+2s-\cos r-\cos q}\nn
   & = &\frac{1}{\sinh\mu}\,[\eps P(r)-\eta e^{-\mu} P_0(r)]. 
\end{eqnarray} 
With 
\begin{equation} \cosh\mu=2+2s-\cos r, 
\end{equation} 
it follows that 
\begin{equation}  
  P_0(r)=\frac{\eps}{\sinh\mu+\eta e^{-\mu}}P(r). 
\end{equation} 
Eliminating $P_0$ now yields 
\begin{eqnarray}  
  \lefteqn{P(r,s)=\Big(s+(1-\gamma)(1-\cos r) }\nonumber\\
  & & 
    +\half\eps[\cos r-\frac{\eta}{\sinh\mu+\eta e^{-\mu}}]-i\va\sin r\Big)^{-1}
  \label{soletad=2}. 
\end{eqnarray} 
For $\eta=\half$ the $\eta$-dependent term becomes $e^{-\mu}$, so then 
the previous situation is recovered. For $\eta=1$ one can check that 
$P(q,r,s)=P_0(r,s)$, showing that no motors reach the fluid (total 
sticking). 
 
For small parameters one has 
\begin{eqnarray}  
  P(r,s)&=&\left(s+\half(1-\gamma-\half\eps)r^2 
    +\half\eps\frac{1-\eta}{\eta}\mu-i\va r\right)^{-1} \nn
   & = & \left(s+\epseff\sqrt{s}-i\va r\right)^{-1}. 
\end{eqnarray} 
So the only effect is the effective detaching probability 
\begin{equation}  
  \eps\to\epseff=\eps\frac{1-\eta}{\eta}. 
\end{equation} 
 
For the probability to be in the fluid no subtraction as in 
(\ref{p2ass=}) is needed.  We have immediately 
\begin{eqnarray}  
  \lefteqn{P(q,r,s) =\frac{2\eps}{4s+4-2\cos q-2\cos r}}\nn 
   & & {}\times\frac{\eta e^{-\mu}+\sinh\mu-\eta\cos q}{\eta e^{-\mu}+\sinh\mu}P(r,s)\nn 
   &\approx & \frac{4\epseff\sqrt{s}}{(q^2+4s)\,(s+\epseff\sqrt{s}-i\va r)}. 
\end{eqnarray} 
 
We can now look for the enhancement of the speed on the tubule; it 
was factor $\pi/2$ in Eq. (\ref{venhd=2}). Let us assume that $\eta$ 
is small, so that there is a large time domain, where we may neglect 
it.  Let us thus set 
\begin{equation}  
  s=\eta^2\sigma,\qquad t=\frac{\tau}{\eta^2},\qquad {\rm and}\qquad
  \eps=\eta^2\tilde\epsilon. 
\end{equation}  
Then we get 
\begin{eqnarray}  
  N_0(s) & = & \frac{1}{\eta^2}\,\frac{1+2\sqrt{\sigma}} 
  {\sigma(1+2\sqrt{\sigma})+\tilde\epsilon\sqrt{\sigma}} \\
  {\rm and}\qquad N_1(s) & = & \frac{\va}{\eta^4}\,\left(\frac{1+2\sqrt{\sigma}} 
    {\sigma(1+2\sqrt{\sigma})+\tilde\epsilon\sqrt{\sigma}}\right)^2. 
\end{eqnarray} 
There are two domains: 
 
1) $\tau\ll 1$, $ \sigma\gg 1$.  Here 
$N_0(t)=e^{-\half\tilde\epsilon\sigma}=e^{-\half\eps t}$, $N_1(t)=\va 
t\,e^{-\half\eps t}$.  Thus $\overline{n}_0(t)=\va t$ and $v(t)=\va$. 
This is still a sharp profile. Although particles have detached, the 
remaining ones go firmly with the bare speed $\va$. 
 
2) $\tau\gg 1$, $\sigma\ll 1$. This we discussed already in 
(\ref{venhd=2}).  The relation $v(t)=\half\pi\va N_0(t)$ just says 
that $v(t)\ll \va$.

\subsection{Variable sticking probability} 
 
Let us now include a variable sticking probability in our model: If a 
motor reaches the filament, it rebinds to it with a probability 
$\pi_{\rm ad}$, while it is reflected from the filament with 
probability $1-\pi_{\rm ad}$. Such a behavior can be due to steric 
constraints, if e.g. a motor with attached bead diffuses close to the 
filament, but with the bead between the motor and the filament.  In 
the long time regime the introduction of this additional parameter is 
expected to reduce to the probability that a motor is bound to the 
filament and thus the effective time-dependent velocity by a factor 
$\pi_{\rm ad}$. This has been confirmed by simulations for the case of 
random walks in open compartments \cite{Lipowsky__Nieuwenhuizen2001}. 
In this section we show analytically that this is indeed the case. 
 
\subsubsection{The two-dimensional case} 
 
Let us begin with the simpler case $d=2$.  To include the sticking 
probability, the master equations for $m=0,\pm 1$ have to be modified: 
On the lines $m=\pm 1$ the rate for hopping to $m=0$, i.e. to the 
filament is $\pi_{\rm ad}/4$, while there is a rate $(1-\pi_{\rm 
  ad})/4$ not to jump, equivalently a motor on these lines attempts to 
hop to the filament with the usual rate $1/4$, but the jump is 
rejected with probability $1-\pi_{\rm ad}$. The modified master 
equations are 
\begin{widetext}
\begin{equation}\label{master_eq_d2_piad_m0} 
  P_{n,0}(t+1)=\frac{\pi_{\rm ad}}{4} P_{n,1}+\frac{\pi_{\rm ad}}{4} P_{n,-1}+(1-\gamma-\half\epsilon-\half\delta)P_{n-1,0}+\frac{\delta}{2} P_{n+1,0}+\gamma P_{n,0} 
\end{equation} 
\begin{equation} 
  P_{n,\pm 1}(t+1)=\oq P_{n+1,\pm 1}+\oq P_{n-1,\pm 1}+\oq P_{n,\pm 2}+\frac{\epsilon}{4} P_{n,0}+\frac{1-\pi_{\rm ad}}{4} P_{n,\pm 1}. 
\end{equation} 
The equivalent of Eq.\ (\ref{p1a=}) now contains also a term with 
$P_1(r,s)$, the Fourier--Laplace transform of the probability 
distribution $P_{n,1}(t)$ along the lines with $m=\pm 1$ adjacent to 
the filament line: 
\begin{eqnarray} 
  \lefteqn{(1+s-\half \cos q-\half\cos r)P(q,r,s)  =  1+ \frac{\pi_{\rm ad}-1}{2}(1-\cos q)P_1(r,s)}\nonumber\\ 
  & & {}+ [\gamma+(\frac{3-2\epsilon-2\delta}{4}-\gamma)e^{ir}-\frac{1-2\delta}{4}e^{-ir}-\frac{1-\epsilon}{2}\cos q]\, P_\bd(r,s).   
\end{eqnarray} 
$P_\bd (r,s)$ and $P_1(r,s)$ are related via the Fourier--Laplace 
transform of Eq.\ (\ref{master_eq_d2_piad_m0}): 
\begin{equation} 
P_1(r,s)=\frac{2}{\pi_{\rm ad}}\left([1+s-(1-\gamma-\frac{\epsilon}{2}-\frac{\delta}{2})e^{ir}-\frac{\delta}{2}e^{-ir}-\gamma]P_\bd (r,s)-1\right). 
\end{equation} 
Using this expression for $P_1(r,s)$, we can proceed in the same way 
as above and obtain 
\begin{equation} 
P_\bd (r,s)=\frac{1+\frac{1-\pi_{\rm ad}}{\pi_{\rm ad}}(1-e^{-\mu})}{[s+(1-\gamma)(1-\cos r)+\frac{\epsilon}{2}\cos r -i\va\sin r ][1+\frac{1-\pi_{\rm ad}}{\pi_{\rm ad}}(1-e^{-\mu})]-\frac{\epsilon}{2}e^{-\mu}}. 
\end{equation} 
\end{widetext}
For small $r$ and $s$, this leads to the asymptotic relation 
\begin{equation} 
 P_\bd (r,s)\approx \frac{1}{s-i\va r+\frac{\epsilon}{\pi_{\rm ad}}\sqrt{s}}, 
\end{equation} 
which has exactly the form of Eq.\ (\ref{p0app}), but with an 
effective detachment rate $\epsilon/\pi_{\rm ad}$.  Doing the 
analogous calculations for the unbound motors we find 
\begin{equation} 
  P_\ub(q,r,s)\approx  
  \frac{4\epsilon\sqrt{s}}{(s-i\va r+\frac{\epsilon}{\pi_{\rm ad}}\sqrt{s})(q^2+4s)}, 
\end{equation}  
which corresponds to Eq.\ (\ref{p2ass=}) again with the effective 
detachment rate $\epsilon/\pi_{\rm ad}$. Hence in the long time regime 
the only effect of the sticking probability $\pi_{\rm ad}$ is a 
rescaling of the detachment rate. Thus the probability for a motor to 
be bound to the filament for large times decays as $N_0(t)\approx 
\pi_{\rm ad}/(\sqrt{\pi}\epsilon\ t^{1/2})$ and the average 
displacement grows as $\sim(\pi_{\rm ad}/\epsilon) \sqrt{t}$, i.e. 
both quantities are reduced by a factor $\pi_{\rm ad}$ as expected 
from the scaling approach \cite{Lipowsky__Nieuwenhuizen2001}.

\subsubsection{The three-dimensional case} 
 
For $d=3$ the calculation is completely analogous. The sticking 
probability is introduced in the same way as above: A motor at a 
neighboring site of the filament attempts to jump to the filament 
with rate $1/6$ as usual, but the attempt is only successful with 
probability $\pi_{\rm ad}$, so that the motor remains at its site with 
probability $(1-\pi_{\rm ad})/6$. Then the probability distributions  
$P(\qb,r,s)$, $P_1(r,s)$ and 
$P_\bd (r,s)$ are related via 
\begin{widetext}
\begin{equation} 
 P_1(r,s)=\frac{3}{2\pi_{\rm ad}}\left([1+s-(1-\gamma-\frac{2\epsilon}{3}-\frac{\delta}{2})e^{ir}- \frac{\delta}{2}e^{-ir}-\gamma]P_\bd (r,s)-1\right) 
\end{equation} 
and 
\begin{eqnarray} 
P(\qb,r,s) & = & \frac{ 3+[3\gamma+\half(5-6\gamma-4\epsilon-3\delta)e^{ir}-\half(1-3\delta)\,e^{-ir}-(1-\epsilon)(\cos q_1+\cos q_2)]\, P_{\rm b}(r,s)}{3+3s-\cos r-\cos q_1-\cos q_2}\nonumber\\ 
 & & +\frac{(1-\pi_{\rm ad})[2-\cos q_1-\cos q_2]P_1(r,s)}{3+3s-\cos r-\cos q_1-\cos q_2}. 
\end{eqnarray} 
\end{widetext}
From these two equations we obtain a rather complicated expression for 
$P_\bd (r,s)$, which, in the limit of small $s$ and $r$, can be reduced to 
\begin{eqnarray} 
  P_\bd (r,s) & \approx & \frac{1}{s-i\va r+\tilde J(s)}\nonumber\\
  {\rm with}\qquad\tilde J(s) & = & \frac{\epsilon}{3\pi_{\rm ad}I(r=0,s)},
\end{eqnarray} 
where $I(r,s)$ is the integral (\ref{integral_d3}). For the unbound 
motors we find 
\begin{equation} 
  P_\ub(\qb,r,s)\approx \frac{\tilde J(s)}{(s-i\va r+\tilde J(s))[s+\frac{q_1^2}{6}+\frac{q_2^2}{6}]}. 
\end{equation} 
Both equations differ from those without the parameter $\pi_{\rm ad}$, 
i.e.\ from the case $\pi_{\rm ad}=1$, only by a rescaling of the 
detachment rate, just as in the two-dimensional case discussed above. 
Therefore in this case too, the long time displacement and the 
probability to be bound to the filament are reduced by a factor 
$\pi_{\rm ad}$, as these quantities are proportional to 
$\epsilon^{-1}$.

\section{Summary and conclusions} 
 
In summary, we have calculated various transport properties arising
from the random walks of molecular motors. Over large length scales
($\gg 1\mu$m), molecular motors perform random walks which consist of
alternating sequences of directed movements along filaments and
non-directed Brownian motion in the surrounding fluid. Here, we have
described these walks as random walks on a cubic lattice and derived
analytical solutions for the cases of a single filament in two or
three dimensions using Fourier--Laplace transforms. We have obtained
closed expressions for the probability distributions of bound and
unbound motors and their moments, which can be evaluated numerically
for all times. The asymptotic behavior at small and large times was
obtained fully analytically. In this way, we derived the fraction of
bound motors, the average position and dispersion, as well as
effective velocities and diffusion coefficients. All these results
were found to be in excellent agreement with results from Monte Carlo
simulations.
 
The random walks of molecular motors exhibit anomalous drift behavior.
In two dimensions the average position of both the bound and unbound
motors grows as $\sim\sqrt{t}$ at large times $t$, while in three
dimensions the displacements grows only logarithmically.  In addition,
diffusion parallel to the filament is strongly enhanced. In the 
two-dimensional case, the diffusion coefficient has an anomalously high
value, which is of the order $(v_\bd/\eps)^2$ where $\epsilon$ denoted
the small detachment probability, see (\ref{Dparline}). In the
three-dimensional case, there are large logarithmic corrections to the
usual diffusion behavior, again of the order $(v_\bd/\eps)^2$, see
(\ref{largeTimeDiff_d=3}).
 
Finally let us emphasize that similar behavior is also obtained for
random walks in confined geometries which have effectively the same
dimensionality \cite{Lipowsky__Nieuwenhuizen2001}. These geometries
are accessible to {\it in vitro} experiments. In addition, unbinding
of motors from filaments and rebinding to them might also be important
for the design of nanotechnological devices using molecular motors as
transport systems, which has been proposed by several groups
\cite{Limberis_Stewart2000,Stracke__Unger2000,Hess__Vogel2001}.

\acknowledgments 

Th.M.N.\ expresses his gratitude for hospitality at
the Max Planck Institute in Golm, where a major part of his work was
done.

\appendix 
\section{Tauberian theorems} 
\label{sec:app:TaubThms} 

The Tauberian theorems allow one to obtain the asymptotic behavior of 
a function $f(t)$ at large times $t$ from the small $s$ behavior of 
its Laplace transform $f(s)={\cal L}[f(t)]$, see, e.g., 
\cite{Feller1971}. The following inverse Laplace transforms ${\cal 
  L}^{-1}[f(s)]$ are used for the random walks of molecular motors:
\begin{widetext} 
\begin{eqnarray}\label{Tauber2}  
  {\cal L}^{-1}[s^{-\alpha}] &\approx &\frac{\al  t^{\al-1}}{\Gamma(1+\al)}\nn  
  {\cal L}^{-1}[-s^{-\alpha}\ln s]&\approx &\frac{t^{\al-1}}{\Gamma(1+\al)}[1+\al \ln t-\al\psi(1+\al)]\nn 
  {\cal L}^{-1}[s^{-\alpha}\ln^2 s]&\approx &\frac{ t^{\al-1}}{\Gamma(1+\al)} [\al \ln^2 t+2\ln t-2\al\ln t \psi(1+\al)-2\psi(1+\al)+\al\psi^2(1+\al)]\nn 
  {\cal L}^{-1}[-\ln^3 s]&\approx &\frac{1}{t}\, [3 (\ln t+\gamma_E)^2-\frac{\pi^2}{3}] . 
\end{eqnarray}
In these expressions, $\Gamma$ is the Gamma function, 
$\Gamma(z)=\int_0^\infty t^{z-1} e^{-t} \d t$, and $\psi$ the Psi 
function defined by $\psi(z)=\d [\ln\Gamma(z)]/\d z$, and 
$\gamma_E\simeq 0.577215$ is Euler's constant. 
\end{widetext}


 
\end{document}